\documentclass[aps,pre,twocolumn,groupedaddress,showpacs,amsfonts,10pt,%
tightenlines,floatfix]{revtex4-1}

\usepackage[utf8]{inputenc}
\usepackage{graphicx}
\usepackage{amsmath,amssymb}
\usepackage{upgreek}
\usepackage[usenames,dvipsnames]{xcolor}
\usepackage{braket}
\usepackage{cancel}

\newcommand{\abs}[1]{\ensuremath{\left\vert#1\right\vert}}
\graphicspath{{./AbbildungenNeu/}}

\begin{document}

\title{Buckling of thermally fluctuating spherical shells: 
   Parameter renormalization and  thermally activated barrier crossing}

\author{Lorenz Baumgarten}
\email[]{lorenz.baumgarten@tu-dortmund.de}

\author{Jan Kierfeld}
\email[]{jan.kierfeld@tu-dortmund.de}

\affiliation{Physics Department, TU Dortmund University, 
44221 Dortmund, Germany}

\date{\today}

\begin{abstract}
We study the influence of thermal fluctuations on the buckling behavior of
thin elastic capsules with spherical rest shape. 
Above a critical uniform pressure, an elastic capsule becomes mechanically
unstable and spontaneously 
buckles into a shape with an axisymmetric dimple. 
Thermal fluctuations affect the buckling instability by two mechanisms.
On the one hand, 
thermal fluctuations can renormalize the capsule's elastic properties
and its pressure because of anharmonic couplings between normal 
displacement modes of different wavelengths. 
This effectively  lowers
 its critical buckling pressure
[Ko\u{s}mrlj and Nelson, Phys.\ Rev.\ X {\bf 7},  011002 (2017)]. 
 On the other hand, buckled shapes are energetically 
favorable already at pressures below the  classical  buckling 
pressure. At these pressures, however,  buckling requires 
to overcome an energy barrier, which 
only vanishes at the critical buckling pressure. 
In the presence of thermal fluctuations the capsule can spontaneously
overcome an  energy barrier of the order of the thermal energy 
 by thermal activation 
already at pressures below the critical buckling pressure. 
We revisit parameter renormalization by thermal fluctuations and 
formulate a buckling criterion based on scale-dependent 
renormalized parameters 
to obtain a  temperature-dependent critical buckling pressure.  
Then we quantify the pressure-dependent 
energy barrier for buckling below the critical buckling
pressure using numerical energy minimization and analytical arguments.  
This allows us to obtain the
temperature-dependent critical pressure for buckling by thermal activation
over this energy barrier.  
Remarkably, both parameter renormalization and thermal activation 
lead to the same parameter dependence of the 
critical buckling pressure on temperature, capsule radius and  thickness,
and Young's modulus.
Finally, we study the combined effect of
parameter renormalization and thermal activation
by using renormalized parameters for the 
energy barrier in thermal activation
to obtain our final result for the temperature-dependent critical pressure,
which is significantly
 below the results if only  parameter renormalization 
or only thermal activation is considered. 
\end{abstract}


\maketitle

\section{Introduction}

Elastic capsules are thin-walled elastic shells enclosing  a fluid
medium.
 Bending energy penalizes deviations in curvature from 
a  specific spontaneous curvature,  and two-dimensional 
elastic energy penalizes stretching and shear 
deformations of the quasi-two-dimensional solid shell with 
respect to a  reference or rest shape, in which the 
capsule is stress free. 
 
On the microscale,  there are many  biological examples 
of elastic capsules such as 
 red blood cells \cite{RBC_Book,RBC_ElCaps}, virus capsids
\cite{Buenemann2007, Virus_Capsids2}, and 
pollen grains \cite{Pollen_Grains}. Microcapsules 
 can also be artificially produced  by various methods, e.g., as hollow
polymer or polyelectrolyte capsules \cite{Artificial1, Artificial2,
  Artificial3}.
Both biological and artificial microcapsules are 
typically used to encapsulate, transport, and eventually release 
 a substance. 
Artificial elastic capsules are used as container and delivery 
systems in numerous applications, such as food technology \cite{Food},
 cosmetics \cite{Cosmetics}, chemical industry,
and pharmacy \cite{Drugs1, Drugs2}.
Also, all macroscopic elastic shells (such as beach balls, egg shells, etc.) 
are described by the same continuum  shell elasticity as microcapsules.

Often the capsule's reference  shape is a sphere (of  radius $R_0$).
A notable exception are red blood cells, where the reference
shape is an oblate spheroid \cite{RBC_Book}.
If the capsule  material can be viewed as a thin shell of thickness
 $h$ ($\ll R_0$)  made from 
an isotropic and homogeneous elastic material, the  shell 
 has a bending modulus $\kappa \propto h^3$ but a  two-dimensional 
Young's modulus $Y\propto h$ \cite{LandauLifshitz,Libai1998}. 
Therefore, bending deformations
are energetically preferred over  stretching
 or shear deformations for thin shells. 
Very thin elastic capsules therefore exhibit a peculiar 
deformation behavior. 
 In the extreme case of an infinitely thin shell, only isometric deformations,
i.e., deformations that preserve the metric and, thus, the Gaussian 
curvature and the stretching and shear energies, 
are possible. 
There is, however,  no smooth and isometric deformation of 
 spheres or ellipsoids, which makes elastic capsules quite 
resistant to pressure or volume decrease,
creating a ``geometry-induced rigidity'' in thin capsules
\cite{Lazarus2012,Vella2012}, which is also employed on the macroscale
for mechanical stability of dome-like structures.

This rigidity makes 
 elastic capsules  stable under  a uniform external pressure, and,
for small pressures, they retain their spherical shape. 
Above a critical  pressure, however, a buckling  instability 
occurs, where the spherical shape becomes mechanically unstable 
and a dimple is finally formed \cite{LandauLifshitz,Timoshenko1961}.
Understanding the buckling instability is both important from a 
structural mechanics perspective, as it is relevant for the mechanical 
stability of macroscopic spherical shells, and in the context 
of microcapsules, which can buckle, for example, by osmotic pressure
\cite{Gao2001,Fery2004,Sacanna2011,Datta2012}.
Following Pogorelov \cite{Pogorelov}, the dimple can be viewed as an 
approximative inverted spherical cap whose sharp edge at the rim 
is rounded to avoid infinite bending energies. 
Such a rounded spherical cap is, therefore, an approximative 
isometry of the spherical rest shape, which avoids large 
stretching energies. 

For ideal spherical shells the classical buckling pressure $p_{c}$
has been known for more than 100 years \cite{Zoelly}.
For a shell with rest radius $R_0$, bending rigidity $\kappa$, 
and two-dimensional (2D) Young's modulus $Y$, 
one finds \cite{Timoshenko1961,Ventsel}
\begin{equation}
  p_{c}  = 4 \frac{\sqrt{Y\kappa}}{R_0^2} =
  4\frac{Eh^2}{R_0^2\sqrt{12(1-\nu^2)}}  = 4 \frac{Y}{R_0}\gamma^{-1/2}.
\label{eq:pcb}
\end{equation} 
The second equality applies for thin shells of thickness $h$ 
made from an isotropic  elastic material with bulk Young modulus $E$ 
and Poisson ratio $\nu$, where $\kappa = Eh^3/12(1-\nu^2)$ and $Y = Eh$
\cite{Ventsel}. We also introduced 
 the F\"oppl--von K\'arm\'an number 
\begin{equation}
  \gamma \equiv \frac{YR_0^2}{\kappa} =
  12(1-\nu^2)\left(\frac{R_0}{h}\right)^{2},
\label{eq:FvK}
\end{equation} 
which is an inverse dimensionless bending rigidity. 
 The ideal 
critical pressure $p_c$ is, however, not reached in experiments on macroscopic 
shells, because imperfections reduce the buckling pressure significantly. 
Such imperfections have been discussed by Hutchinson \cite{Hutchinson1967}
and Koiter  \cite{Koiter1969}  in the form of additional 
quenched normal displacements.

At $p_c$ (or a $p_c$ reduced by imperfections) buckling occurs as 
an instability  with respect to oscillatory 
normal displacements, the shortest possible  wavelength  of which is 
$\lambda_c =  2\pi R_0 \gamma^{-1/4} = 2 l_\text{el}$ \cite{Hutchinson1967}, 
which sets an important elastic  length scale 
\begin{equation}
l_\text{el} \equiv \pi \left(\frac {\kappa R_0^2}{Y}\right)^{1/4} 
 = \frac{12^{1/4}\pi}{(1-\nu^2)^{1/4}} \sqrt{R_0h}  = \pi R_0 \gamma^{-1/4} 
\label{eq:lel}
\end{equation}
in the buckling problem. 
For thin shells with $\gamma \gg 1$,  the 
buckling wavelength is small compared to  
 the shell radius (but large compared to shell thickness). 
 There are many unstable modes with this wavelength, and the buckling
 instability  results in a hexagonal lattice of dimples of size $\sim
 \lambda_c$ on the sphere, as was shown theoretically 
\cite{Hutchinson1967,Koiter1969} and also 
 by  experiments \cite{Carlson} and 
numerical simulations \cite{Paulose2013}.
This buckling pattern is unstable with respect to growth of one 
of the dimples on the expense of the others, finally resulting 
in a single axisymmetric dimple. For a fixed mechanical 
pressure $p\ge p_c$, the dimple will actually snap through and grow 
until opposite sides are in contact, whereas for osmotic pressure
control or even volume control, a stable dimple shape is reached before 
opposite sides come into contact \cite{Knoche2011,Knoche2014o}.
The dimple can, however, assume a polygonal shape 
in a {\it secondary} buckling transition \cite{Quilliet2012,
Knoche2014,Knoche2014a}.

Interestingly, the buckled state with a single axisymmetric dimple 
becomes {\it energetically} favorable already at a much lower pressure 
$p_{c1} \ll p_c$, which is sometimes
also called {\it Maxwell pressure} because it can be obtained from a 
Maxwell construction of equal energies 
\cite{Knoche2011,Knoche2014o,Hutchinson2017b,Evkin2017}. 
For $p_{c} >p>p_{c1}$, the 
axisymmetrically buckled configuration with a single dimple has a 
 lower energy as compared to a spherical shape, but the 
spherical shape remains a local energy minimum, which is protected
by an {\it energy barrier} from buckling \cite{Knoche2011,Knoche2014o}.
\footnote{The classical buckling pressure $p_c$ was denoted by 
$p_{\rm cb}$ and the   critical pressure for equal energies 
by $p_c$  in Refs.\ \cite{Knoche2011,Knoche2014o} 
  or $p_{\rm 1st}$  in  Refs.\ \cite{Knoche2014a,Knoche2014}.
}
For mechanical pressure control, a parameter dependence 
\begin{equation}
p_{c1} \sim  p_{c} \gamma^{-1/4}\sim \frac{Y}{R_0}\gamma^{-3/4}
   \sim \frac{Eh^{5/2}}{R_0^{5/2}} 
\label{eq:pc1}
\end{equation} 
 has been found \cite{Knoche2014o}.
The results of Ref.\ \cite{Knoche2011} show that 
there is also a  critical {\it unbuckling pressure}
 $p_\text{cu}$, below which no stable buckled shape exists.
In Ref.\ \cite{Knoche}, $p_\text{cu}  \sim 3p_{c1}/4$ has been found; i.e.,
$p_\text{cu}$ has the same parameter dependence as $p_{c1}$.
The unbuckling pressure  also  corresponds to  the 
minimum pressure on the pressure-volume  relation of the buckled 
branch, for which 
 the same  parameter has been found in 
Refs.\ \cite{Evkin2001,Evkin2017}.

\begin{figure}
\begin{center}
 \includegraphics[width=1\linewidth]{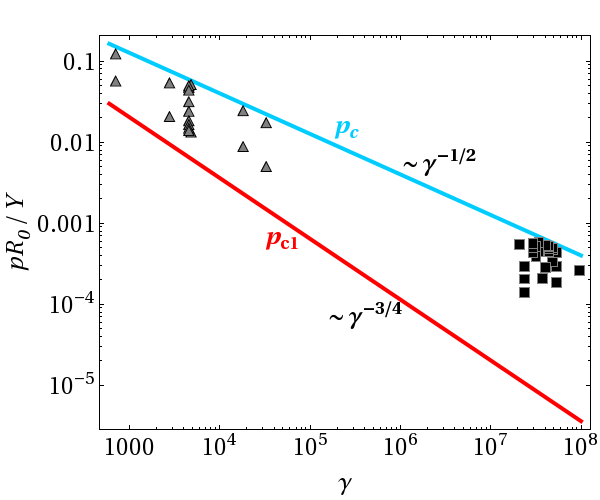}
\caption{
Classical  buckling  pressure $p_{c}\propto \gamma^{-1/2}$ 
(dark red line) and critical pressure
  $p_{c1}\propto \gamma^{-3/4}$ (light blue line), 
where buckling is already energetically favorable according to
Ref.\ \cite{Knoche2014o}. 
Experimental results (black squares, from Ref.\ \cite{Carlson}) 
and finite-temperature Monte Carlo simulations 
(gray triangles, from Ref.\ \cite{Paulose2012}) for the
buckling pressure lie between $p_{c}$ and $p_{c1}$.
}
\label{fig:pc_knoche}
\end{center}
\end{figure}

As a result, there is a rather wide pressure window  $p_{c} >p>p_{c1}$,
where buckling is energetically possible but must be induced 
by imperfections  or other external perturbations because an 
energy barrier has to be overcome. 
This energy barrier has  been subject of a number of recent 
studies both for spherical 
\cite{Marthelot2017,Evkin2017,Hutchinson2017,Hutchinson2017b,Thompson2017}
and cylindrical \cite{Virot2017,Thompson2017} shells and 
will  be quantified in this paper for spherical shells by numerical 
calculations.
One important  cause of perturbation
 to be studied within this work and particularly 
relevant for thin shells or capsules are  thermal fluctuations.
Thin shells or two-dimensional elastic materials deform easily by bending 
and can therefore exhibit pronounced thermal shape fluctuations 
at room temperature \cite{Ahmadpoor2017}.
Thermal fluctuations 
could give rise to thermal activation over the buckling 
energy barrier. 
Figure \ref{fig:pc_knoche}
shows that both experimental values for macroscopic shells 
as well as simulation results for thermally fluctuating 
shells always lie within the pressure window  $p_{c} >p>p_{c1}$,
where buckling is energetically allowed.

This suggests that thermal fluctuations of a shell or capsule 
can have two effects on the 
buckling transition: 
(i) They induce shape fluctuations similar to the 
imperfections studied in Refs.\ \cite{Hutchinson1967,Koiter1969}. 
Therefore, they 
should {\it reduce} the classical buckling pressure $p_c$ 
where a spontaneous 
mechanical instability sets in. 
(ii) They give rise to thermal activation over the buckling energy barrier 
in the pressure window between $p_{c}$ and $p_{c1}$, which leads to a
 further reduction of the apparent buckling pressure $p_c$ in the 
presence of thermal fluctuations.

The effect (i) has already been studied analytically and numerically 
in Refs.\ \cite{Paulose2012, Kosmrlj2017}. Using renormalization group (RG)
calculations, it has been established  
that thermal shape fluctuations of a shell renormalize its 
Young modulus downward and bending rigidity upward
 and generate an effective compressive
 pressure \cite{Kosmrlj2017}. 
As a result, the classical buckling pressure is {\it decreased} by 
thermal shape fluctuations, i.e., the shell effectively
softens and becomes increasingly unstable with temperature,
which has also been observed in 
Monte-Carlo simulations of thermally fluctuating shells \cite{Paulose2012}.
Using the same RG treatment as Ref.\ \cite{Kosmrlj2017}
but different buckling criteria, we will arrive at essentially 
the same conclusions. 

The effect (ii) of thermal activation over the buckling energy barrier 
has not been considered so far and is the main 
subject of the present paper. 
First we will quantify the pressure-dependent 
buckling energy barrier $F_\text{B}$ using numerical 
energy minimization. 
We find a scaling behavior 
$F_\text{B}(p) \sim YR_0^2 \gamma^{-3/2}  f_p\left({p}/{p_c}\right)
\sim   (Eh^4/R_0) f_p\left({p}/{p_c}\right)$ of the buckling energy barrier 
with  a scaling function $f_p(x) \sim (1-x)^2x^{-3}$.
Then 
we  consider the sole effect 
of thermal activation without parameter renormalization and 
calculate an apparent 
buckling pressure from arguing that, at a finite temperature, 
energy barriers of the size of the thermal energy $k_BT$ ($k_B$ is the 
Boltzmann constant and $T$ temperature) can be 
overcome quasi-spontaneously on experimental time scales by 
thermal activation. 
Finally,  we will study the combined effect of  (i) parameter renormalization 
and (ii) thermal activation and find a further reduction of the 
apparent critical buckling pressure both below the results 
if only  parameter renormalization 
or only thermal activation is considered.

We can compare our results to existing numerical Monte-Carlo 
simulation  results \cite{Paulose2012}, which show that, 
for $T>0$, the critical buckling pressure $p_c(T)$ is 
only weakly decreasing with temperature for $k_BT< \kappa/\sqrt{\gamma}$
but exhibits a pronounced decrease for $k_BT> \kappa/\sqrt{\gamma}$.
The simulation data could be collapsed onto a curve 
$p_c(T)/p_c(0) = f(k_BT\sqrt{\gamma}/\kappa)$, where $f(x)$ is a 
scaling function. We will obtain the same scaling of the critical 
buckling pressure with $k_BT\sqrt{\gamma}/\kappa$ 
{\it both} for (i) parameter renormalization 
{\it and} (ii) thermal activation because 
$F_B(p) \sim \kappa /\sqrt{\gamma} f_p\left({p}/{p_c}\right)$.

\section{Buckling and parameter renormalization}

We start by recapitulating the RG approach of 
Refs.\  \cite{Paulose2012, Kosmrlj2017}  leading to the 
RG transformation for the scale-dependent elastic moduli and a 
scale-dependent pressure. 
Based on these RG equations we discuss several buckling 
criteria, which are slightly different from those used 
in Ref.\ \cite{Kosmrlj2017} but give rise to very 
similar temperature-dependent  buckling pressures.

\subsection{Elastic energy and thermal fluctuations}

\begin{figure}
\begin{center}
 \includegraphics[width=1\linewidth]{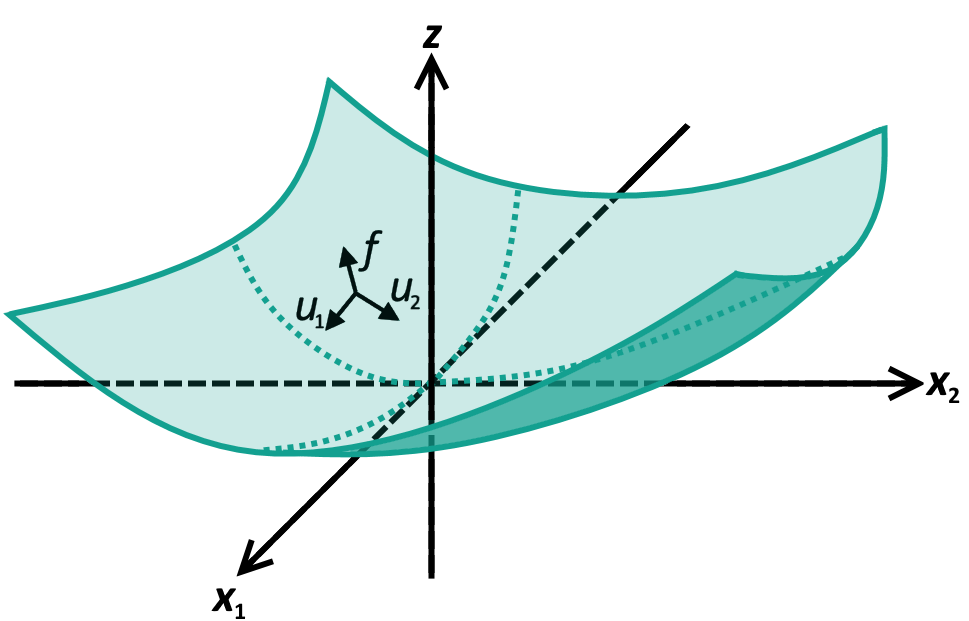}
\caption{
Coordinate system used in the shallow shell theory.
}
\label{fig:ShShCoords}
\end{center}
\end{figure}

The elastic energies of a spherical 
shell or  capsule can be calculated using 
shallow shell theory \cite{Koiter1969}, which 
 is accurate for weakly curved  shells
with $h \ll R_0$ or, 
 alternatively, at large  F\"oppl--von K\'arm\'an numbers 
$\gamma = {YR_0^2}/\kappa \gg 1$ [see Eq.\ (\ref{eq:FvK})].
 In shallow shell
theory, the undeformed reference state of a 
 nearly flat section of the sphere is described by a height function
$Z({\bf x})$, where the Cartesian coordinates ${\bf x} = (x_1,x_2)$ 
define a tangent plane that
touches the sphere at the origin  as shown 
in  Fig.\ \ref{fig:ShShCoords}.
For a spherical shell the reference state
 has constant mean curvature $1/R_0$.

The reference state is deformed   by  tangential displacements
${\bf u}({\bf x})$ [${\bf u}=(u_1,u_2)$]
and  normal displacements $f({\bf x})$ ($f>0$ shall correspond 
to inward displacements); see Fig.\ \ref{fig:ShShCoords}. 
For small displacements of the spherical reference shape,
we obtain a strain tensor \cite{LandauLifshitz}
\begin{equation}
 u_{ij} = 
   \frac{1}{2} (\partial_iu_j + \partial_j u_i + \partial_i f \partial_j f) 
    - \delta_{ij}\frac{f}{R_0} 
\label{eq:uij}
\end{equation}
and a corresponding in-plane elastic energy 
\begin{equation}
 E_S = \frac{1}{2} \int dS (2\mu u_{ij}^2 + \lambda u_{kk}^2),
\label{eq:FS}
\end{equation}
with the Lam\'e coefficients $\mu$ and $\lambda$ 
and the area element $dS$. 

For the bending energy, we can use  the Helfrich form \cite{Helfrich}
(assuming $\nu=1$ for bending \cite{Knoche2011}),
\begin{equation}
 E_b = \frac{\kappa}{2} \int dS (2H-2H_0)^2,
\label{Helfrich}
\end{equation}
 where $\kappa$ is  
the bending rigidity, $H$ is the mean curvature,  
and $H_0=1/R_0$ is the spontaneous
mean curvature of the undeformed sphere.
For a shallow section of the shell, the mean
curvature can be written as
\begin{equation}
 2H = \nabla ^2 (Z+f) = \frac{2}{R_0} + \nabla^2 f,~~ \nonumber
\label{eq:H}
\end{equation}
 where $\nabla^2 = \partial_{11} + \partial_{22}$. 

Adding the mechanical work  by an external
pressure $p$ (with the sign convention that 
$p>0$ is a compressive pressure), 
the total enthalpy is
\begin{align}
 F &= E_S + E_b  - p\int dS f \nonumber \\ 
  &= \int d^2 x \left[ \frac{\kappa}{2} \left( \nabla^2 f\right)^2 
    + \mu u_{ij}^2 + \frac \lambda 2 u_{kk}^2 - p f\right].
\label{eq:Ftot}
\end{align}
The strain tensor (\ref{eq:uij}) contains 
terms $\partial_i f\partial_j f$ and the coupling $f/R_0$ of normal
displacements to the  Gaussian background 
curvature, both of which give rise to anharmonicities in the normal 
displacement field $f$ in the enthalpy. 
The latter source of anharmonicities is characteristic for 
bent shells and vanishes in the limit $R_0\to \infty$ of a planar 
plate or membrane. 
The free enthalpy $G$ is obtained by integrating   out all thermal
displacement fluctuations, 
$G = -k_B T \ln \left[ \int \mathcal{D} {\bf u}
    \int df e^{-F/k_B T}\right]$, which can not be achieved 
analytically in explicit  form because of the anharmonicities. 
In a RG calculation, this integration  is performed iteratively, starting 
from small scales, and contributions from anharmonicities are absorbed 
into renormalized scale-dependent elastic moduli $\kappa_R$ and $Y_R$ and 
a scale-dependent renormalized pressure $p_R$.

To this end, 
the normal displacement is separated  into a uniform part $f_0$ caused 
by the homogeneous pressure $p$ and a
nonuniform part $\tilde f({\bf x})$ solely due to thermal fluctuations,
\begin{equation}
 f({\bf x}) = f_0 + \tilde f({\bf x}) = f_0 
    + \sum_{{\bf q}\ne 0} \tilde f({\bf q})
   e^{-i{\bf q}\cdot {\bf x}},
 \label{eq:f(x)}
\end{equation}
with Fourier components 
 $ \tilde f({\bf q}) 
    = A^{-1} \int d^2 x f({\bf x})  e^{i{\bf q}\cdot {\bf x}}$,
where $A$ is the  area in the ${\bf x}$ plane.
The  enthalpy $F$
is  split into a harmonic and an anharmonic part. Only the uniform normal
displacement contributes to the pressure term, and only the nonuniform term
contributes to the nonlinear part of the strain tensor. Therefore,
the enthalpy 
is harmonic in $u_i({\bf x})$ and $f_0$, and these  fields are eliminated
by defining an effective enthalpy \cite{Nelson_Crumpling}
\begin{align}
 F_\text{eff}[\tilde f] = -k_B T \ln \left[ \int \mathcal{D} {\bf u}
    \int df_0 e^{-F/k_B T}\right].
\label{eq:F_eff0}
\end{align}

Integrating out the  phonon
 fields $u_i$ and  homogeneous modes $f_0$, we obtain \cite{Paulose2012}
\begin{align}
 F_\text{eff}[\tilde f]
&= \int d^2 x \left( \frac{1}{2} \left[\kappa \left( 
    \nabla^2 \tilde f\right)^2 - \frac {pR_0} 2|\nabla \tilde f|^2 
    + \frac {Y\tilde f^2}{R_0^2}\right] \right.\nonumber \\
&\left.~~~+ \frac{Y}{8} \left[P_{ij}^T 
   (\partial_i \tilde f)(\partial_j \tilde f)\right]^2 
   - \frac {Y\tilde f}{2R_0}\left[P_{ij}^T 
  (\partial_i \tilde f)(\partial_j \tilde f)\right]\right),
\label{eq:F}
\end{align}
(we sum over double indices)
where $P_{ij}^T = \delta_{ij} - \partial_i\partial_j/\nabla^2$ is the
transverse projection operator and $Y=  4\mu (\mu+\lambda)/(2\mu+ \lambda)$
 is 2D Young's modulus.  The last cubic term is characteristic for 
bent shells and vanishes in the limit $R_0\to \infty$ of a planar 
plate or membrane. Appearance of a cubic term simply reflects a missing mirror
symmetry for bent shells. 

The corresponding Fourier transformed
correlation function $G({\bf q}) \equiv \braket{|\tilde f ({\bf q}) |^2}$ 
is in harmonic approximation  \cite{Paulose2012}
\begin{equation}
 G_0({\bf q}) = \frac {k_B T}{A\left[\kappa q^4 
      - \frac{1}{2} pR_0q^2 + \frac{Y}{R_0^2}\right]}.
\label{eq:Gff}
\end{equation}
In the RG approach 
 anharmonic contributions are absorbed into 
scale-dependent 
renormalized parameters $\kappa_R(q)$, $p_R(q)$, and $Y_R(q)$  that 
replace $\kappa$, $p$, and $Y$ in (\ref{eq:Gff}) such that 
the full  
correlation function shall finally be obtained in the form  
\begin{equation}
 G({\bf q}) = \frac {k_B T}{A\left[\kappa_R(q) q^4 
      - \frac{1}{2} p_R(q)R_0q^2 + \frac{Y_R(q)}{R_0^2}\right]}.
\label{eq:GR}
\end{equation}

\subsection{Renormalization group transformation}
\label{RGTh}

Now, we shortly recapitulate the derivation of the  scale-dependent
parameters  $\kappa_R(q)$, $p_R(q)$, and $Y_R(q)$
by   Ko\u{s}mrlj and Nelson \cite{Kosmrlj2017}.
In the RG transformation,
fluctuations in the normal displacement field $\tilde{f}$ are integrated out 
up to the  length scale $l = a b$ or down to a wave number 
$q = \pi/l = \Lambda/b$, where $b=e^s$ is the  scale factor 
of the RG transformation and $a = \pi/\Lambda$ is a microscopic cutoff 
scale (e.g., the shell thickness). 
Ko\u{s}mrlj and Nelson have shown that the exact choice of
$a$ is irrelevant as long as $a\ll l_\text{el}$, with the 
{\it elastic length scale} (\ref{eq:lel}), which is the 
length scale of initial dimples at the classical buckling instability,
 and 
$a \ll l_\text{th}$, with  the {\it thermal length scale}
$l_\text{th} \equiv  \sqrt{{16\pi^3\kappa^2}/{3k_B T Y}}$,
which is the length scale above which  
thermal fluctuations become relevant for flat plates. 
After integrating out, the original cutoff is re-established 
by rescaling 
lengths and fields according to
\begin{equation}
   {\bf x}= b{\bf x}'~,~~ {\bf q}= b^{-1}{\bf q}'
 ~,~~\tilde f({\bf q}) = b^{\zeta_f}\tilde f'({\bf q}'),
\label{rescale}
\end{equation}
with a field rescaling exponent $\zeta_f$.

In a momentum shell RG approach, the normal
  displacements $\tilde f$  are separated into slow modes
  $\tilde f_<({\bf x}) = \sum_{\abs {\bf q} < \Lambda/b} e^{i{\bf q}\cdot{\bf x}}
   \tilde f ({\bf q})$ and fast modes 
  $\tilde f_>({\bf x}) = \sum_{\abs {\bf q} > \Lambda/b} e^{i{\bf q}\cdot{\bf x}} 
   \tilde f ({\bf q})$ containing modes with  wave numbers smaller and larger 
 than $\Lambda/b$, respectively. 
We integrate over modes $\tilde f_>$
 in the momentum shell 
 $\Lambda/b < \abs {\bf q} < \Lambda$ to obtain an 
effective enthalpy for the slow modes $\tilde f_<$,
\begin{equation*}
 F'_\text{eff}[\tilde f_<]= -k_B T 
  \ln \left( \int \mathcal D [\tilde f_>({\bf x})] e^{-F_\text{eff}/k_B
    T}\right).
\end{equation*}
Then we rescale lengths and fields according to Eq.\ (\ref{rescale}). 
The effective enthalpy for slow modes retains 
its form \eqref{eq:F} by this 
 change of scale by a factor  $b$ if  new renormalized 
elastic parameters  $\kappa'(s)$, $Y'(s)$,
and $p'(s)$ are introduced. 
Their RG flow  for an infinitesimal change of scale $b\approx 1-ds$ 
is described by $\beta$-functions
\begin{subequations}
\begin{align}
 \beta_\kappa &= \frac {d\kappa'}{ds} 
  = 2(\zeta_f -1)\kappa' +
 \frac {3k_B T Y' \Lambda^2}{16\pi D} \nonumber \\ 
  &~~~- \frac {3k_B T Y'^2\Lambda^2}{8\pi R_0'^2D^2}
  \left[ {\frac{11}{12}}  + \frac{I_{\kappa 1}} {D^2} 
   + \frac {I_{\kappa 2}}{D^4}\right],
\label{beta_kappa}\\
\beta_Y &= \frac {d Y'}{ds} 
   = 2\zeta_f Y' - \frac {3k_B T Y'^2\Lambda^6}{32\pi D^2},\\
\beta_p &= \frac{dp'}{ds} 
   = (2\zeta_f +1)p' + \frac {3k_B T Y'^2 \Lambda^4}{4\pi R_0'^3 D^2}
     \left[1 + \frac {I_p} {D^2}\right],
\\
\beta_R &= \frac {dR_0'}{ds} = -R_0',
\end{align}
\label{BetaFunctions}
\end{subequations}
where the denominator
 \begin{align}
  D &\equiv
     \kappa'\Lambda^4 - \frac {p'R_0'\Lambda^2} 2 + \frac {Y'}{R_0'^2}
\label{eq:D}
\end{align}
was introduced. 
The terms $I_{\kappa 1}$,~$I_{\kappa 2}$,  and $I_p$ are given 
in Eq.\ (\ref{eq:Is}) in Appendix \ref{diagramsApp}. 
The function $\beta_\kappa$ (\ref{beta_kappa}) seemingly 
differs in two terms 
from the results of Ref.\ \cite{Kosmrlj2017} but is actually 
identical.
For a consistent renormalization of  the three- and four-point vertices,
we choose  $\zeta_f = 1$.

Finally, the scale-dependent 
 quantities $\kappa_R(q)$, $p_R(q)$, and $Y_R(q)$,
in terms of which the full correlation function (\ref{eq:GR}) 
can be written, are obtained by undoing the rescaling operation;
this gives (using $\zeta_f=1$)
\begin{subequations}
 \begin{align}
  \kappa_R(q) &= \kappa'(s) e^{(2-2\zeta_f)s} = \kappa'(s),\\
  Y_R(q) &= Y'(s) e^{-2\zeta_fs} = Y'(s) e^{-2s},\\
  p_R(q) &= p'(s) e^{(-1-2\zeta_f)s} = p'(s)e^{-3s},~\text{and}\\
  R_{0,R}(q) &= R_0'(s) e^s  = R_0,
 \end{align}
\label{eq:Rquantity}
\end{subequations}
where $s=s(q)$ is given by $e^s = \Lambda/q$. 
These parameters arise by simply integrating out fluctuations 
up to the scale $\ell=\pi/q=a e^s$, i.e., modes with 
wave numbers $>q$, {\it without} subsequent rescaling 
and obey the  RG equations (\ref{BetaFunctionsR}) 
given in Appendix \ref{diagramsApp}.
For any other quantity, the index ``R'' also indicates that it is 
a scale-dependent, i.e., renormalized but unrescaled quantity.

\subsection{Buckling criteria}
\label{Buckling_Criteria}

The classical buckling instability at the 
pressure $p_{c}  = 4 {\sqrt{Y\kappa}}/{R_0^2}$
[see Eq.\ (\ref{eq:pcb})] manifests in the presence of thermal 
fluctuations
as the smallest pressure where the denominator 
$\kappa q^4  - \frac{1}{2} pR_0q^2 + {Y}/{R_0^2}$ of the 
correlation function 
$G_0({\bf q})$ in harmonic approximation from Eq.\ (\ref{eq:Gff})  
can become zero, which happens at the wave number
$q_\text{el}
 = (p_cR_0/4\kappa )^{1/2} = (Y/\kappa R_0^2)^{1/4} = \pi/l_\text{el}$
[see Eq.\ (\ref{eq:lel})], 
which indeed corresponds to the wavelength $\lambda_c$ of the 
classical buckling instability. 
At this  point, the shell becomes energetically unstable 
with respect to small fluctuations in the 
radial deformation mode $\tilde f(q_c)$, 
which initiates the buckling transition.

In the presence of thermal fluctuations, elastic constants and pressure 
are renormalized according to (\ref{BetaFunctions}) and become 
dependent on the length scale  $\ell \sim 1/q$ up to which 
fluctuations have been integrated out:
$\kappa_R(q)$ grows, $Y_R(q)$ decreases, 
and also the pressure $p_R(q)$ 
grows for decreasing $q$ or increasing length scales $\ell$. 
At $T>0$ buckling happens  analogously to 
the mechanical  $T=0$ case
 if the denominator of the correlation function $G(q)$ 
(the  two-point vertex function), 
\begin{equation}
  D_R(q) \equiv k_BT/AG(q) = \kappa_R (q) q^4 - \frac{1}{2}
     p_R(q) R_0 q^2 + \frac {Y_R(q)}{R_0^2},
\label{eq:DRq}
\end{equation}
is renormalized to zero for a certain wave number $q=q^*$, starting 
from the initial ``bare'' values 
$p_R(\Lambda)=p$ (and  $Y_R(\Lambda)=Y$, $\kappa_R(\Lambda)=\kappa$) 
at    $q=\Lambda$. Then the correlation
function diverges, and the radial deformation mode $\tilde f(q^*)$ becomes
unstable in the effective harmonic theory with renormalized parameters 
and initiates the  buckling transition. 
The smallest initial value $p$ for which $D_R(q)=0$ is reached 
for  some $q=q^*$ is the critical buckling  pressure $p_c(T)$. 
The corresponding  unstable wave number 
 $q^*$ replaces the  elastic wave number
$q_\text{el} =  (Y/\kappa R_0^2)^{1/4} = \pi/l_\text{el}$ for the 
$T=0$  buckling instability,  $q^*(T\!=\!0)=q_\text{el}$. 
It is important to note that $q^*$ is not 
identical with the renormalized value $q_\text{el,R} = 
 [Y_R(q_\text{el,R})/\kappa_R(q_\text{el,R}) R_0^2]^{1/4}$ of 
$q_\text{el}$, because  Eq.\ (\ref{eq:bc0}) also contains contributions 
from the derivatives $\kappa_R'(q)$, $p_R'(q)$, and $Y_R'(q)$ 
 [see also Eq.\  (\ref{BetaFunctionsR}) in  Appendix \ref{diagramsApp}].

Because $p_c(T)$ is the smallest initial  $p$ 
for which $D_R(q)$ acquires a zero at $q=q^*$, 
$D_R(q)$ also has to have a minimum (or saddle) at $q=q^*$. 
Two cases  have to be distinguished: This minimum can be
 in the interior of the interval of possible $q$ values, 
$\pi/R_0 < q^* < \Lambda$, or it can be 
a boundary minimum at the smallest $q$ value $q^*=\pi/R_0$. 
For an interior minimum or saddle with 
 $\pi/R_0 < q^* < \Lambda$,
both $D_R(q^*)=0$ and $\partial_q D_R(q^*)=0$ are fulfilled
at the  critical buckling pressure, which 
leads to 
\begin{align}
0 =D_R(q) &=\left[\kappa_R (q) q^4 
  - \frac{1}{2} p_R (q) R_0 q^2 + \frac {Y_R(q)}{R_0^2}\right] 
  \nonumber \\
 0 = \partial_q D_R(q)&= 4\kappa_R (q) q^3 - p_R (q) R_0 q  
\nonumber\\
   &~~~+ \left[\kappa_R'(q) q^4 
  - \frac{1}{2} p_R'(q) R_0 q^2 + \frac {Y_R'(q)}{R_0^2}\right].
\label{eq:bc0}
\end{align}
These two equations determine
both the renormalized critical buckling pressure 
$p_\text{c,R}= p_R(q^*)$ 
and the  corresponding unstable wave number $q^*$.
If Eqs.\ (\ref{eq:bc0})  lead to a  $q^*<\pi/R_0$, 
 the actual  minimum is at the boundary 
value $q^*=\pi/R_0$, and buckling happens at the longest scale, 
$l^* = {\pi}/{q^*}=R_0$. Then,   the single equation 
$D_R(\pi/R_0)=0$  determines  
the renormalized critical buckling pressure $p_\text{c,R}= p_R(\pi/R_0)$. 
For both cases, 
the  buckling pressure $p_{c}(T)$ is obtained 
as the bare initial  value $p = p_c(T)$ of the RG transformation 
which has to be chosen to reach $p_R(q)=p_\text{c,R}$ at the 
corresponding buckling wave number $q=q^*$.

In order to determine $p_c(T)$ numerically, 
instead of  solving Eqs.\ (\ref{eq:bc0}), we 
start
at a small bare initial $p$ and follow the 
RG flow from $q=\Lambda$ down to the smallest 
$q=\pi/R_0$. If $D_R(q)=0$ occurs,
buckling happens. The smallest initial $p$ for which this happens
is the  buckling pressure $p_{c}(T)$, and the wave number 
$q$ for which this  happens is the unstable wave number
$q^*$.

In Ref.\  \cite{Kosmrlj2017}, a slightly different 
buckling criterion was employed, namely 
that there exists a $q=q^*$
 where the  renormalized 
external pressure $p_R(q)$ reaches the 
renormalized critical buckling pressure: $p_R(q^*)=p_{c,R}(q^*) = 
{4\sqrt{\kappa_R(q^*) Y_R(q^*)}}/{R_0^2}$. This is similar, but 
not  equivalent, to our  criterion $D_R(q^*)=0$
in conjunction with $\partial_q D_R(q^*)=0$ for a local minimum, 
because Eq.\ (\ref{eq:bc0}) also contains contributions 
from $\kappa_R'(q)$, $p_R'(q)$, and $Y_R'(q)$, which are neglected 
if $p_{c,R}(q^*) = {4\sqrt{\kappa_R(q^*) Y_R(q^*)}}/{R_0^2}$ is used.

The criterion of Ref.\ \cite{Kosmrlj2017}
can also be interpreted by considering 
the vertex function
$D_R(q^*,q) = \kappa_R (q^*) q^4 - 
     p_R(q^*) R_0 q^2/2 + {Y_R(q^*)}/{R_0^2}$,
which is obtained by integrating out all fluctuations with 
$\Lambda > q > q^*$,
and which governs the remaining long wavelength 
fluctuations with wave numbers $q$ in $q^* > q > \pi/R_0$.
The vertex function $D_R(q^*,q)$ actually has an 
instability if $p_R(q^*)=p_{c,R}(q^*) = 
{4\sqrt{\kappa_R(q^*) Y_R(q^*)}}/{R_0^2}$, but the unstable 
wave number is  the renormalized elastic wave number 
$q_\text{el,R} =  [Y_R(q^*)/\kappa_R(q^*) R_0^2]^{1/4}$, which differs 
from $q^*$ in general. 
If  $q_\text{el,R}>q^*$, this unstable mode is even no longer 
accessible to the shell, as 
it has already been integrated out.
As long as differences between $q^*$ and $q_\text{el,R}$ are small, 
the criterion $p_R(q^*)=p_{c,R}(q^*)$ should give 
comparable results to our criterion  $D_R(q^*)=0$.

\subsection{Critical pressure from parameter renormalization}
\label{RG-Results}

Ko\u{s}mrlj and Nelson have shown that the results
 of the RG-transformation are
solely dependent on a dimensionless temperature 
\begin{equation}
 \bar{T} \equiv \frac{k_BT\sqrt{\gamma}}{\kappa} 
\sim \frac{k_BT R_0}{Eh^4}
    \sim \frac{l_\text{el}^2}{l_\text{th}^2}.
\label{eq:barT}
\end{equation} 
Because our buckling criterion $D_R(q^*)=0$ operates 
on renormalized parameters, the critical buckling pressure 
$p_c(T)$ should also only depend on $\bar{T}$.

\begin{figure}
\begin{center}
\includegraphics[width=1\linewidth]{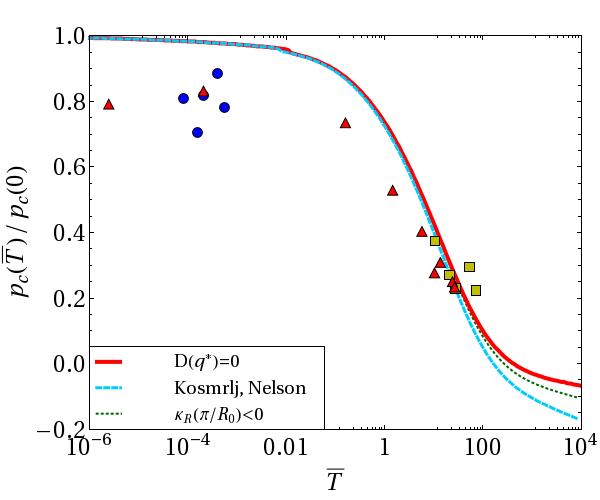}
\caption{
 Critical buckling pressure $p_c(T)$  as a function of the dimensionless
  temperature $\bar{T}= {k_BT\sqrt{\gamma}}/{\kappa} $ 
   according to the  buckling criterion  $D_R(q^*)=0$ 
 (dark red solid line) and 
  according to the criterion
$p_R(q^*)=p_{c,R}(q^*)$ from Ko\u{s}mrlj and Nelson \cite{Kosmrlj2017}
 (light blue dashed line).
   The  criterion $p_R(q^*)=p_{c,R}(q^*)$
  gives slightly smaller values of $p_c(T)$
but both criteria lead to $p_c(T)<0$ for high temperatures,
signaling that fluctuating shells can  spontaneously buckle even without 
external compressive pressure.
We also show the pressure where $\kappa$ first renormalizes to zero 
at the largest scale,  $\kappa_R(\pi/R_0)=0$ (green dotted line). 
This pressure is always
below the critical $p_c(T)$  from the 
 buckling criterion  $D_R(q^*)=0$, signaling that there are already
fluctuations with $\kappa_R<0$ present before buckling. 
  Data points show numerical results from Ref.\ \cite{Paulose2012}. 
}
\label{fig:pcs}
\end{center}
\end{figure}

\begin{figure}
 \includegraphics[width=1\linewidth]{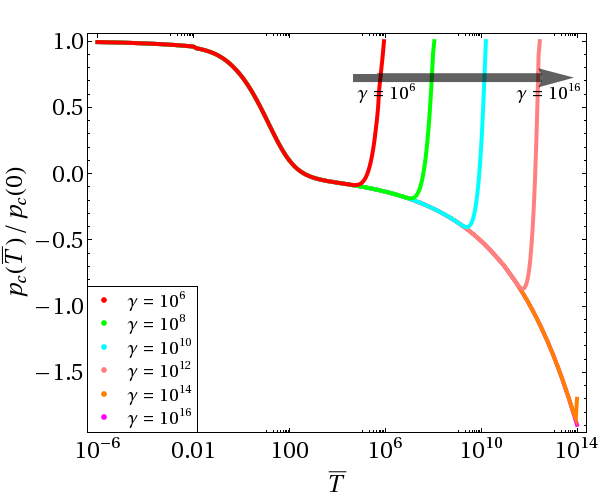}
\caption{Non-universality of the critical 
buckling pressure $p_\text{c}(\bar T)$ 
for different F\"oppl--von K\'arm\'an numbers $\gamma$ (see arrow).
For large values of $\gamma$, the non-universality 
is shifted to large dimensionless temperatures $\bar{T}$.
}
\label{fig:non-universal}
\end{figure}

\begin{figure}
\begin{center}
\includegraphics[width=1\linewidth]{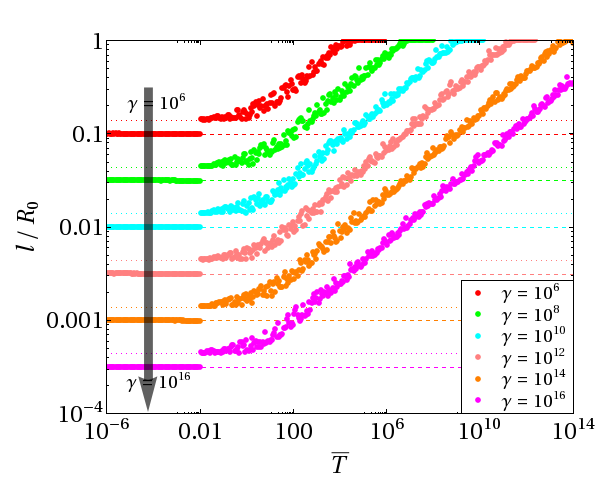}
\caption{Length scale $l^*$ 
on which buckling occurs at the critical buckling
  pressure as a function of the dimensionless temperature $\bar{T}$
for different F\"oppl--von K\'arm\'an numbers $\gamma$ 
(see arrow).
   Dashed lines show
  the $T=0$  elastic length scale $l_\text{el}/R_0$  [Eq.\ (\ref{eq:lel})] 
and dotted lines show
  $1.4\,l_\text{el}/R_0$ (see text).
}
\label{fig:lengthscale}
\end{center}
\end{figure}

Figure \ref{fig:pcs} shows the critical buckling
pressure $p_c(T)$ as a function of the dimensionless temperature $\bar{T}$ 
for our  buckling criterion $D_R(q^*)=0$ and 
the criterion $p_R(q^*)=p_{c,R}(q^*)$ used in Ref.\ \cite{Kosmrlj2017}. 
All RG flows have been
calculated using an eighth-order Runge-Kutta method.
First we can confirm that, for both buckling criteria, $p_c(T)$ 
only depends on the dimensionless temperature $\bar{T}$  for 
a wide range of  F\"oppl--von K\'arm\'an numbers $\gamma \gg 10^3$.
This is also corroborated by the  scaling collapse of numerical simulation 
results in Ref.\ \cite{Paulose2012} (see also data points in 
Fig.\ \ref{fig:pcs}).
Only for large $\bar{T}$ do results  become non-universal; 
see Fig.\ \ref{fig:non-universal}.
This happens if  the  length scale 
$l^* = {\pi}/q^*$  on which
buckling occurs according to the criterion $D_R(q^*)=0$ 
reaches the largest accessible length scale $R_0$ such 
that $p_c(T)/p_c(0)$ becomes also $\gamma$ dependent, 
as can be seen from comparing  Figs.\  \ref{fig:non-universal}
and Fig.\ \ref{fig:lengthscale}.
For  small F\"oppl--von K\'arm\'an numbers 
$\gamma < 10^3$, already the elastic length scale $l_\text{el}$
on which buckling occurs at $T=0$ is comparable to the radius $R_0$,
and the classical buckling pressure $p_c$ from Eq.\ (\ref{eq:pcb}) 
is no longer applicable.

For small temperature $\bar{T} <1$, both criteria give practically 
identical results. 
The results are also in rough agreement with the numerical Monte-Carlo 
simulation  results from Ref.\ \cite{Paulose2012}.
At larger temperatures $\bar{T} \gg 1$,  
the  criterion $p_R(q^*)=p_{c,R}(q^*)$ 
  gives slightly smaller values of $p_c(T)$, i.e., 
  slightly 
 underestimates the buckling stability. 
Both criteria give rise  to $p_c(T)<0$ for high temperatures,
signaling that fluctuating shells can  spontaneously buckle even without 
external compressive pressure \cite{Kosmrlj2017}.

In Fig.\ \ref{fig:pcs}, we also show the initial pressure  for which
 $\kappa_R(q)$ first renormalizes to zero.
This always happens 
at the largest scale $q=\pi/R_0$ because of the shape of the 
RG flow of $\kappa_R(q)$: The renormalized $\kappa_R(q)$ 
is first increasing with $q$ decreasing from its starting value $q=\Lambda$ 
but, if the renormalized denominator $D_R(q)$ becomes small, 
exhibits a maximum followed by a sharp decrease to  $\kappa_R<0$.
Therefore, negative values $\kappa_R<0$ are always attained at the 
largest scales, 
 and $\kappa_R(q)=0$ is first fulfilled for $q=\pi/R_0$ if 
starting from small initial pressure. 
 The (blue) line of initial pressures $p$ for which 
 $\kappa_R(\pi/R_0)=0$  is, except for very small 
dimensionless temperatures $\bar T \lesssim 10^{-4}$,
{\it below} the critical $p_c(T)$  from the 
 buckling criterion  $D_R(q^*)=0$ (red line)  in Fig.\ \ref{fig:pcs}.
 This shows that there are always
unstable bending fluctuations with $\kappa_R<0$ already 
present before buckling.  These are, however, still stabilized by 
the last positive term in the vertex function $D_R(q)$ in Eq.\ 
(\ref{eq:DRq}), which originates from the unavoidable stretching 
deformation that comes with any bending deformation of a sphere. 
This effect cannot be captured  by the  criterion 
$p_R(q^*)=p_{c,R}(q^*)\propto \sqrt{\kappa_R(q^*)}$ because 
it will always be fulfilled {\it before} $\kappa_R=0$ is reached. 
Therefore, the (green) line for the initial pressure for which
 $\kappa_R(q)$ first renormalizes to zero lies {\it between}
the  buckling pressures $p_c(T)$ obtained with the two criteria 
$D_R(q^*)=0$ and $p_R(q^*)=p_{c,R}(q^*)$ in   Fig.\ \ref{fig:pcs},
and the critical buckling pressures from the criterion $D_R(q^*)=0$ 
always lies {\it above} the critical pressure from the 
criterion $p_R(q^*)=p_{c,R}(q^*)$.

 Figure  \ref{fig:lengthscale} shows the length scale 
$l^* = {\pi}/q^*$  on which
buckling occurs according to the criterion $D_R(q^*)=0$ 
as a 
function of the dimensionless temperature $\bar{T}$ and 
for different   F\"oppl--von K\'arm\'an numbers $\gamma$.
 For large temperatures, the buckling length
scale approaches the radius $R_0$, whereas for small temperatures it
approaches the classical $T=0$  elastic length scale $l_\text{el}$ from 
Eq.\ (\ref{eq:lel}).
This means that renormalization of elastic constants and 
pressure  effectively transforms 
buckling  into a long wavelength instability at higher temperatures. 
For small F\"oppl--von K\'arm\'an numbers $\gamma$,
the buckling length
scale approaches the radius $R_0$ already 
for smaller dimensionless temperatures $\bar{T}$.
If the buckling length $l^*$ approaches $R_0$, results 
for the critical buckling pressure $p_c(T)$ become 
non-universal as discussed above (see Fig.\ \ref{fig:non-universal}). 

\begin{figure}
\begin{center}
\includegraphics[width=1\linewidth]{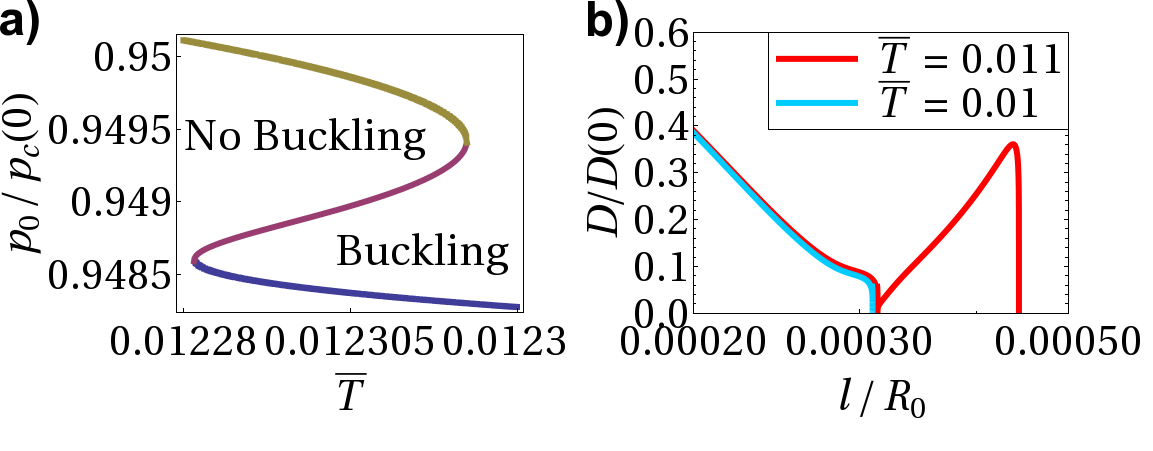}
\caption{
(a)  Critical buckling pressure $p_c(\bar{T})$ in the temperature range 
 $0.01228<\bar{T} <0.01233$.
(b)
$D_R(\ell)$ develops a local minimum as a function of $\ell/R_0$ 
for temperatures below and above the
   jump in the buckling length scale 
 [$\bar{T}=0.01$ (dark red) and $\bar{T}=0.011$ (light blue), respectively]. 
 The jump occurs when the minimum reaches zero, which triggers the        
    buckling criterion $D_R(\ell) = 0$ for $\ell  = l^*$.
}
\label{fig:pTJump}
\end{center}
\end{figure}

The temperature dependence of the length scale $l^*$
in Fig.\ \ref{fig:lengthscale} shows another remarkable feature 
for temperatures around
 $\bar{T} \simeq 0.011$, where it 
abruptly jumps from values slightly below the elastic length scale to
values about  $1.4\,l_\text{el}$, depicted by the dotted lines.
This relation
between the length scale before and after the jump is found 
 for a wide range of F\"oppl--von K\'arm\'an numbers $\gamma$.  
The jump in the buckling length scale can be traced back to the existence of
an additional local minimum of $D_R(q)$ 
for temperatures in the range
$0.01<\bar T < 25$.  Typically, slightly below the critical pressure 
$D_R (q)$ has only one 
local minimum, which develops into a divergence towards negative infinity
at $q^*$  if
the critical buckling pressure is reached.  The jump occurs if 
two minima are present when the critical buckling pressure is 
reached and when the global minimum exchanges 
between both minima right at the buckling 
pressure [see Fig.\ \ref{fig:pTJump}(b)], 
 meaning that there are two  $q^*$ values 
for which $D_R(q)=0$.
Then the buckling length scale  changes discontinuously, whereas 
the  critical buckling
pressure remains continuous.

At slightly higher  temperatures $\bar{T} \simeq0.0123$, 
also the critical buckling pressure exhibits an interesting 
feature, as it  drops  abruptly from 
$p_c(T)/p_c(0)= 0.950$ to $p_c(T)/p_c(0) = 0.949$. 
In this temperature range an increase in the bare pressure $p$ above
the red  line in  Fig.\ \ref{fig:pTJump}(a) first 
leads to buckling; i.e.,  $D_R(q^*)=0$  is fulfilled for a certain $q^*$. 
After increasing $p$  
above the yellow line, however, 
we find that again $D_R(q)>0$ for all $q$. Only  above the blue line,
 a $q^*$ with $D_R(q^*)=0$ exists again.
This leads to an S-shape $p_c(T)$ curve in  
Fig.\ \ref{fig:pTJump}(a). If the temperature is 
increased past the end of the blue line in  Fig.\ \ref{fig:pTJump}(a),
the buckling pressure drops abruptly.

This feature in the $p_c(T)$ curve 
 could be related to the jump in $l^*$ and, thus, to the jump in $q^*$ 
which we discussed 
before. This jump  happens, however, at slightly lower temperatures. 
Both the jumps in $l^*$ and in $p_c(T)$ 
around $\bar{T}\simeq 0.011-0.012$ might be  artifacts
of the approximate RG flow and might  disappear if the 
RG calculation is extended to higher  loop order.

\section{Buckling by thermal activation over the buckling energy barrier}

Now we address the main issue of the paper, namely 
buckling by thermal activation over the 
buckling energy barrier. 
We use the Pogorelov approach and numerical 
simulations to determine the energy barrier associated 
with the formation of a stable axisymmetric dimple
for pressures $p<p_c$, i.e., below the critical buckling pressure. 
If only thermal activation over the energy barrier  
drives the buckling process, we estimate the critical buckling pressure 
using the criterion 
that barriers of order $k_BT$ can be
 overcome quasi-spontaneously on experimental time scales by 
thermal activation. 
Including also renormalization effects of elastic 
parameters into the energy barrier calculation,
we describe the combined effect of  thermal activation and parameter 
renormalization. 
We will also present evidence from numerical simulations that 
 states with multiple dimples are not relevant 
for the energy barrier and, thus, for thermal activation.

\subsection{Elastic energies and  energy barrier in the Pogorelov model}
\label{PogSection}

For the following energetic considerations for indented 
configurations of spherical shells we will not use the 
shallow shell approximation but the general 
form (\ref{eq:FS}) of the stretching energy, the Helfrich form 
 (\ref{Helfrich})  of the bending energy, and the mechanical work 
$E_p=-p\Delta V$, where $\Delta V$ is the volume reduction with respect 
to the initial rest state with $V_0 = 4\pi R_0^3/3$. 
Stretching and bending energy can be brought into dimensionless 
form by measuring energies in units of $YR_0^2$ and lengths 
in units of $R_0$. Using also  $Y=  4\mu (\mu+\lambda)/(2\mu+ \lambda)$
for Young's modulus 
and $\nu = \lambda/(2\mu+\lambda)$ for Poisson's ratio
in the stretching energy, we obtain 
\begin{align}
 \frac {E_S}{YR_0^2} &= \frac{1}{2(1+\nu)}
    \int \frac{d S}{R_0^2}
    \left(u_{ik}^2 + \frac{\nu}{1-\nu} u_{ll}^2\right),
\nonumber\\
\frac {E_b}{YR_0^2} &= \frac{2}{\gamma} \int d S (H-H_0)^2,
\nonumber\\
 \frac{E_p}{YR_0^2} &= -\frac{pR_0}{Y} \frac{\Delta V}{R_0^3}.
\label{eq:Fdim}
\end{align}
We see that the deformation behavior of the sphere only depends on the 
two parameters
$\gamma$ and $\nu$ characterizing the elasticity of the shell 
and the 
dimensionless  pressure $pR_0/Y$ or, alternatively, $p/p_c$ 
(because $p_c = 4(Y/R_0)\gamma^{-1/2}$).

Pogorelov approximated the energy  of an axisymmetric 
 dimple on a sphere by assuming that the
dimple is an approximate isometric deformation of the sphere, i.e.,
 that the dimple is an inverted spherical cap of 
 the undeformed sphere, where the edges of the
dimple are rounded by bending energy. 
Such a   mirror inversion of a spherical cap 
is a suitable starting point to approximate buckled configurations
because it avoids  additional 
stretching strains. Using this approximation, 
Pogorelov 
calculated the  energy of an axisymmetric indentation of 
volume $\Delta V$ on a sphere of initial volume $V_0 = 4\pi R_0^3/3$
as \cite{Pogorelov,Knoche2014o}
\footnote{
Note that Pogorelov used a slightly different form 
of the bending energy in Ref.\ \cite{Pogorelov} which included
a Poisson number $\nu$ to account for an anisotropy of  bending deformations. 
The Helfrich energy corresponds to $\nu=1$. 
This difference has  practically no effect on our results. 
}  
\begin{align}
 U_\text{Pog}(\Delta V,V_0)
 = 
 c_\text{Pog} \frac {YR_0^2}{(1-\nu^2)^{1/4}}\gamma^{-3/4}
 \left(\frac {\Delta V}{V_0}\right)^{3/4},
\label{UPog}
\end{align}
where $\nu$ is Poisson's ratio
and $c_\text{Pog}\approx 15.09$, 
$V_0$ is the initial
volume of the sphere, $V<V_0$ is its volume after indentation, 
and $\Delta V =V_0-V$ is the volume reduction by dimple formation.
Slightly more accurate estimates of the energy
of an axisymmetric dimple are given in Refs.\ \cite{Gomez2016,Evkin2017}
but we will use the Pogorelov estimate (\ref{UPog}) in the following.

The Pogorelov energy (\ref{UPog}) neglects 
that, under pressure $p$, the spherical shell is already 
uniformly compressed  before the
indentation is formed.
Uniform compression to a volume $V_0-\Delta V$ 
 costs an energy
\begin{align}
   U_{\rm sph}(\Delta V) &=4\pi \frac{YR_0^2}{1-\nu} 
   \left[ \left( 1- \frac{\Delta V}{V_0} \right)^{1/3} -1 \right]^2
\nonumber\\
&{\approx}
     \frac{4\pi}{9}   \frac{YR_0^2}{1-\nu}
    \left( \frac{\Delta  V}{V_0}\right)^2,
\label{eq:Usph}
\end{align}
where the last approximation holds for $\Delta V \ll V_0$. 
The equilibrium  volume  follows from 
 $p=dU_\text{sph}/d\Delta V$,
resulting in 
\begin{equation}
  p \approx 
    \frac{2}{3} \frac{1}{1-\nu}  \frac{Y}{R_0} \frac{\Delta  V}{V_0}.
\label{psph}
\end{equation}
During buckling the spherical body relaxes this pre-compression 
but will remain
 compressed to a volume 
$V_0-\Delta V_b$ (with $\Delta V_b<\Delta V$), in addition to forming
 an indentation with volume 
reduction  $\Delta V-\Delta V_b$, such that $\Delta V$ 
is the total volume reduction. 
The remaining pre-compression of the  spherical body costs an energy 
$U_\text{sph}(\Delta V_b)$ as given by Eq.\ (\ref{eq:Usph}),
the formation of the additional indentation an energy 
$U_\text{Pog}(\Delta V-\Delta V_b,V_0-\Delta V_b)$ as given by
 Eq.\ (\ref{UPog}).
The optimal buckled shape at pressure $p$
 is then obtained by minimizing the total enthalpy
\begin{align}
 F(p,\Delta V_b,\Delta V) &= 
      U_{\rm sph}(\Delta V_b) \nonumber\\
   &~~ + U_\text{Pog}(\Delta V-\Delta V_b,V_0-\Delta V_b)
-p\Delta V
\label{min}
\end{align}
with respect to the spherical pre-compression volume deficit 
$\Delta V_b$ {\it and} the  total volume deficit $\Delta V$.
Under volume control, the total energy  $U(\Delta V_b,\Delta V)=
 U_{\rm sph}(\Delta V_b) +
  U_\text{Pog}(\Delta V-\Delta V_b,V_0-\Delta V_b)$ 
is minimized with respect to $\Delta V_b$ only, at
{\it fixed}  total volume deficit $\Delta V$.
We neglect in Eq.\ (\ref{min}) the influence of the homogeneous 
compressional background 
stress associated with the pre-compression by a volume $\Delta V_b$ 
on the Pogorelov energy $U_\text{Pog}$, which  is approximately 
justified as the inverted Pogorelov cap  changes stretching strains
only at the edges of the indentation. 
For $\Delta V_b\ll V_0$,  we can approximate 
$U_\text{Pog}(\Delta V-\Delta V_b,V_0-\Delta V_b)\approx 
U_\text{Pog}(\Delta V-\Delta V_b,V_0)$. 
Then, equilibrium states under volume control 
become equivalent to equilibrium states under pressure control 
with an effective  pressure $p = dU_{\rm sph}(\Delta V_b)/d\Delta V_b$
generated by the compressional stress of the spherical body.

\begin{figure}
\begin{center}
\includegraphics[width=1\linewidth]{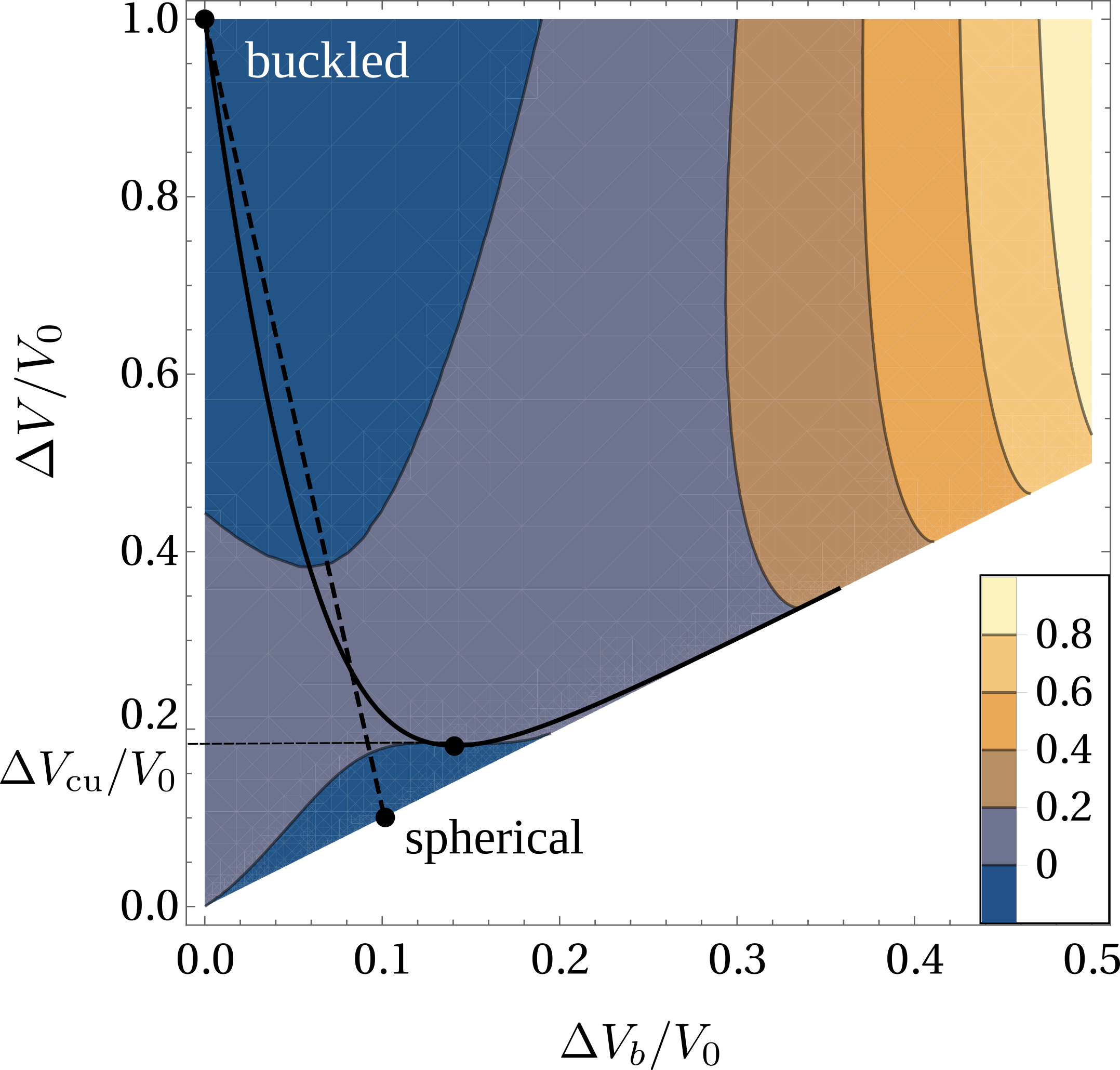}
\caption{
Contour plot of Pogorelov enthalpy $F(p,\Delta V_b,\Delta V)/YR_0^2$ as a 
function of pre-compression volume $\Delta V_b$ of the spherical 
capsule body and total volume deficit $\Delta V$
with $\Delta V_b\le \Delta V$
($\gamma = 100$, $p R_0/Y=0.15>p_{c1}$, $\nu=1/2$, 
$F$ in units of $YR_0^2$). 
Solid black lines represent local minima and maxima as a 
function of $\Delta V_b$, which only exist above the 
critical unbuckling volume deficit $\Delta V_\text{cu}$.
Along the dashed path from the metastable spherical shape 
to the buckled shape, the energy barrier $F_\text{B}$  from Eq.\ 
(\ref{FbarrTh}) has to be overcome.
}
\label{fig:Poglandscape}
\end{center}
\end{figure}

Analyzing the $\Delta V_b$-dependence of $U(\Delta V_b,\Delta V)$
for fixed $\Delta V$
shows  that the compressed spherical shape with  $\Delta V_b=\Delta V$
remains a metastable minimum for {\it all} $\Delta V$ (see
Fig.\ \ref{fig:Poglandscape}). 
This is  an artifact of the Pogorelov approximation, which is not able 
to predict the classical buckling  pressure $p_c$ or the critical 
classical  buckling volume given by   \cite{Knoche2011}
\begin{align}
   \frac{\Delta V_{c}}{V_0}  &= 1-\left[\frac{1}{2} + 
  \sqrt{\frac{1}{4} + 2(1 - \nu)\gamma^{-1/2}}
     \right]^{-3}  
   \nonumber\\
&\approx  6(1-\nu) \gamma^{-1/2}
\label{eq:Vcb}
\end{align}
(the last approximation applies to  $\gamma\gg 1$).
This  is the volume where the compressive pressure 
in a sphere reaches 
the classical buckling pressure  $p_c = 4 (Y/R_0)\gamma^{-1/2}$
[see Eq.\ (\ref{eq:pcb})] according to the 
 pressure-volume relation  (\ref{psph}) 
in the compressed  spherical shape. 

Under volume control, 
the total energy $U(\Delta V_b,\Delta V)$ can 
develop a second local minimum 
as a function of $\Delta V_b$, 
which  corresponds to a buckled state  (see
Fig.\ \ref{fig:Poglandscape}).
This minimum  with respect to $\Delta V_b$ only exists for sufficiently 
large total volume reduction
\begin{align}
   \frac{\Delta V}{V_0}  \ge  \frac{\Delta V_\text{cu}}{V_0} 
  &= \frac{5}{4^{4/5}} \left(\frac{27 c_{\rm Pog}}{32\pi}\right)^{4/5} 
     \frac{(1-\nu)^{4/5}}{(1-\nu^2)^{1/5}} \gamma^{-3/5}
   \nonumber\\
   &= 5.05 \frac{(1-\nu)^{4/5}}{(1-\nu^2)^{1/5}} \gamma^{-3/5},
\label{Vcu}
\end{align}
and we obtain  a  {\it  critical unbuckling  volume deficit}
$\Delta V_\text{cu}$ 
below which the buckled state has to 
 spontaneously  ``unbuckle.''
This lower critical  volume deficit also  corresponds to  the 
minimum volume on the pressure-volume  relation of the buckled 
branch and has also been found in Ref.\ \cite{Evkin2017}.
The critical unbuckling volume deficit shows a different power-law dependence 
$\Delta V_\text{cu}\propto \gamma^{-3/5}$ on the 
F\"oppl--von K\'arm\'an number as compared to the 
classical  buckling volume $\Delta V_{c}\propto \gamma^{-1/2}$ 
  in Eq.\ (\ref{eq:Vcb}).
For $\Delta V>\Delta V_\text{cu}$, minimization with respect to 
$\Delta V_b$ gives the  optimal  pre-compression
\begin{align}
    \frac{\Delta  V_b}{V_0}
  &\approx \frac{27 c_{\rm Pog}}{32\pi}
      \frac{1-\nu}{(1-\nu^2)^{1/4}}  \gamma^{-3/4}
\left(\frac{\Delta V}{V_0}\right)^{-1/4} 
\label{DeltaVpre}
\end{align}
in the buckled state
(for $\Delta V_b \ll \Delta  V$).

Under pressure control, 
the buckled state becomes a local minimum of 
the enthalpy  $F(p,\Delta V_b,\Delta V)$ as a function of both $\Delta V_b$
and $\Delta V$ only for sufficiently 
large pressure $p>p_\text{cu}$, i.e., above a {\it lower critical 
unbuckling pressure}, which is given by the condition that 
the maximally buckled state $\Delta V = V_0$ ($\Delta V_b=0$) 
becomes metastable:
$p_\text{cu} = (\partial U_\text{Pog}/\partial \Delta V)(V_0,V_0)$. 
We find $p_\text{cu} \propto (Y/R_0)\gamma^{-3/4}$, again with a 
different parameter dependence as compared to the classical 
buckling pressure 
$p_{c} \propto (Y/R_0)\gamma^{-1/2}$ [see Eq.\ (\ref{eq:pcb})]. 
The maximally buckled state becomes energetically favorable 
over the spherical state 
for $p>p_{c1}$, i.e., above the Maxwell pressure 
 $p_{c1} = 4p_\text{cu}/3 \propto (Y/R_0)\gamma^{-3/4}$
\cite{Knoche2011,Knoche2014o}.

In the modified Pogorelov enthalpy landscape 
$F(p,\Delta V_b,\Delta V)$ (see Fig.\ \ref{fig:Poglandscape}), the metastable 
spherical state $\Delta V_b=\Delta V$, which exists for 
$p>p_\text{cu}$, is always protected 
by an energy barrier from spontaneous transitions 
into the buckled state. 
This energy barrier can be determined from 
starting at a spherical initial state $\Delta V_b=\Delta V=\Delta V_i$ and 
moving along a path $(\Delta V_b, \Delta V) = (\Delta V_i,\Delta V_i) + 
  v(-A,1)$  in the $\Delta V_b$-$\Delta V$ plane 
into an arbitrary direction $(-A,1)$ with decreasing $\Delta V_b$ ($A>0$)
using  a volume parameter $v\ge 0$. 
Independently of $A$ we find 
an energy barrier 
\begin{align}
F_\text{B}(p) &= a(\nu) YR_0^2 \gamma^{-3/2} \left(\frac {p}
{p_c}\right)^{-3}
   = \tilde{a}(\nu) \frac{Eh^4}{R_0}  \left(\frac {p}{p_c}\right)^{-3}
\label{FbarrTh}
\end{align}
with
\begin{equation*}
a(\nu) \equiv 
 \frac {27^2}{\pi^32^{20}}   
       c_\text{Pog}^4 \frac{1}{1-\nu^2}
\simeq \frac{1.16}{1-\nu^2},
\end{equation*}
which is assumed 
at  a total  indentation volume $(\Delta V-\Delta V_b)/V_0 = 
(1+A)v/V_0 \sim \gamma^{-1}(p/p_c)^{-4}$.
This result is only  valid for  $p/p_c \gg \gamma^{-1/4}$ 
(or $p\gg p_\text{cu}$) such that
$(\Delta V-\Delta V_b)/V_0 \sim \gamma^{-1}(p/p_c)^{-4} \ll 1$ and dimples 
remain  small compared to the total capsule volume. 
The transition state, where this maximum is assumed, is a sphere
with  an energetically unfavorable  ``flattened'' dimple,
which is unstable both with respect to shrinking back to a sphere 
and growing into a fully developed dimple. For a fixed mechanical 
pressure, the fully developed dimple  actually snaps through 
until opposite sides are in contact. Only for a 
volume-dependent osmotic pressure
 or under strict volume control, a stable dimple of finite 
size is possible \cite{Knoche2011,Knoche2014o}.
The  buckling shapes that are assumed around the 
barrier if also multiple dimples are allowed are discussed below 
in Sec.\ \ref{sec:shapes}. The result (\ref{FbarrTh}) 
describes the energy barrier for a single axisymmetric dimple.

It is clear that the Pogorelov model is incorrect for  very large and 
very 
small dimples as the assumption that the dimple can be described as a rounded
 mirror inversion fails in both of these limits. 
For very small dimples, this can be easily seen  by the fact
that the energy barrier  (\ref{FbarrTh}) 
does not  disappear for  $p\ge p_c$.
We can, however, assume that very shallow  dimples of a 
depth $\zeta$ comparable  to the layer thickness  $h$ 
(assuming a thin shell made from an isotropic elastic material), i.e., 
$\zeta \sim h$, can be formed spontaneously (a similar assumption is 
made in Ref.\ \cite{LandauLifshitz} to derive the classical buckling 
pressure $p_c$). Using the fact that the 
dimple opening angle is  $\alpha\sim  \sqrt{\zeta/R_0}$ and $\Delta V \sim 
 \alpha^4 R_0^3 \sim \zeta^2 R_0$ such a small dimple 
of volume $\Delta V/V_0 \sim h^2/R_0^2 \sim \gamma^{-1}$
costs an energy 
$U_\text{Pog}/YR_0^2 \sim 
  \gamma^{-3/4} (h/R_0)^{3/2}\sim \gamma^{-3/2}$ according to Eq.\
 (\ref{UPog}). Comparing it with the barrier scaling (\ref{FbarrTh}),
we conclude that exactly at $p\sim p_c$ the energy barrier 
reduces to this energy and could, thus, be overcome spontaneously. 
The Pogorelov  result (\ref{FbarrTh}) for the energy barrier 
is only a good approximation for $p\ll p_c$ (in practice,
 $p< 0.8\, p_c$, see Fig.\ \ref{fig:Num-Pog} in  Appendix \ref{sec:SE}) 
such that dimples 
are large enough, but  $p/p_c \gg \gamma^{-1/4}$ 
(or $p\gg p_\text{cu}$) such that 
$\Delta V/V_0 \sim \gamma^{-1}(p/p_c)^{-4} \ll 1$ and dimples 
remain  small compared to the total capsule volume. 
If $p$ approaches $p_c$ the buckling energy barrier actually 
vanishes as our simulation results in the next section show.

\subsection{Simulation results for the energy barrier}
\label{Simulation}

In order to investigate the behavior of  the energy 
barrier for single axisymmetric dimples 
also for  $p \le  p_c$, i.e., for small dimples 
more rigorously, we 
use numerical energy minimization with the 
SURFACE EVOLVER \cite{SEExpMath}.
Some details are explained in Appendix \ref{sec:SE}.

Recently, a number of publications addressed the energy barrier 
for axisymmetric dimples on spherical shells 
by  applying  an additional point force 
in order to induce formation of a single dimple, 
where the point force $F$ is applied, and in order to control 
the indentation depth $\zeta$ by the point force 
\cite{Evkin2017,Hutchinson2017,Hutchinson2017b,Evkin2017,Thompson2017}. 
The barrier state corresponds to an indented state with $F=0$ at 
$\zeta = \zeta_B$, which 
is unstable with respect to growth and shrinkage. 
The energy barrier is obtained from the $F(\zeta)$-relation by 
$U_B = \int_0^{\zeta_B} F(\zeta) d\zeta$. 
An additional point force has also been used 
in experiments  \cite{Marthelot2017} in order to 
calculate the energy barrier. 
In our numerical approach we will not prescribe a point force 
but directly constrain the conjugated indentation depth $\zeta$.

\begin{figure}
 \includegraphics[width=1\linewidth]{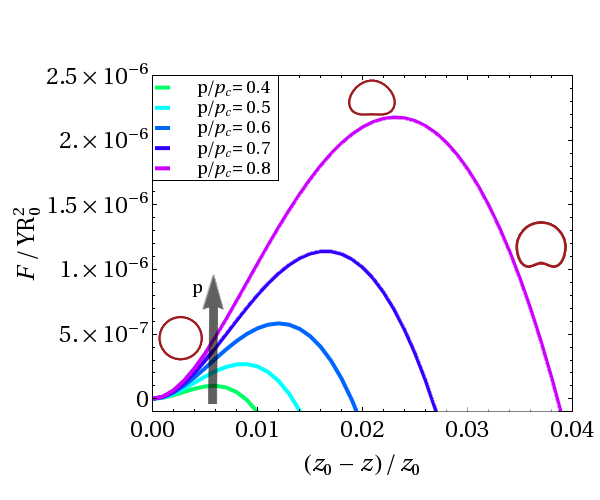}
\caption{
Numerical results for the 
enthalpy as a function of the dimensionless indentation 
depth
$(z-z_0)/z_0 = 2\zeta/R(p)$ for $\gamma=10^5,~\nu=0.3$,
 and for different values of $p/p_\text{c}$ [choosing $F(z=z_0)=0$ for 
an initial vertex distance $z_0=2R(p)$].
The buckling energy barrier is obtained from the enthalpy maximum. 
Representative schematic shapes are shown for
 the spherical initial state, the barrier state 
with a  ``flattened'' dimple, and a  buckled state with a 
well-developed dimple (note that, under pressure control, 
the fully developed dimple corresponding to the buckled energy minimum 
will snap through until opposite sides are in contact). 
The indentations in the schematic 
shapes are exaggerated; the actual dimple indentations are much smaller,
c.f.\ values on the $x$-axis.
}
\label{fig:Fz}
\end{figure}

\begin{figure*}
\begin{center}
\includegraphics[width=1\linewidth]{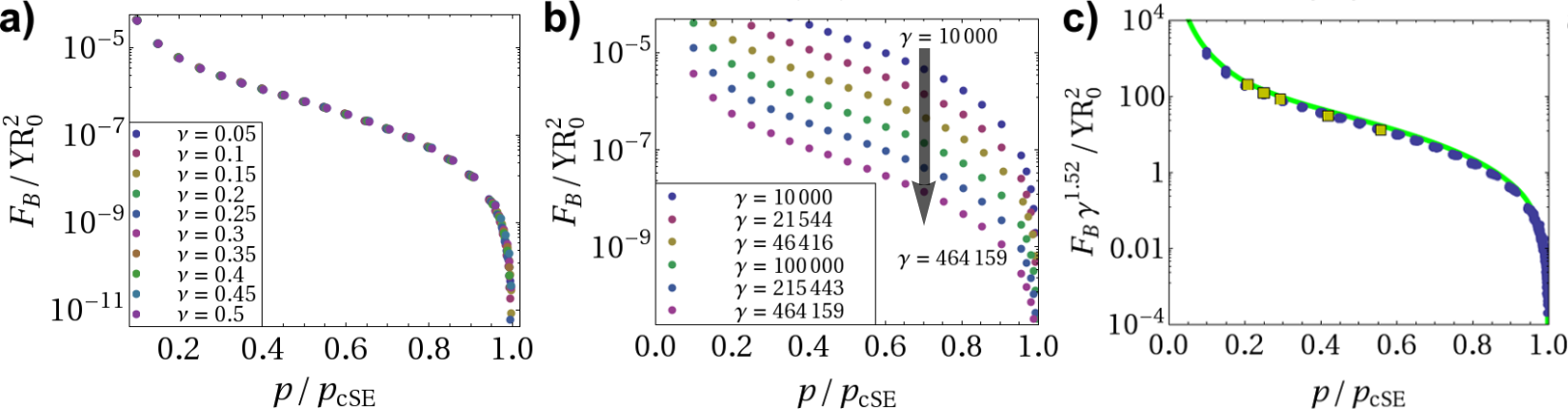}
\caption{Energy barrier as a function of the pressure for
    (a) several values of
  $\nu$ at $\gamma = 10^5$  and (b) several values of $\gamma$ (see
    arrow)  at
  $\nu=0.3$.
 (c) Collapse of the energy barrier for different values of $\gamma$
  (dots). The  numerical approximation $f_\text{p,num}(p/p_\text{c,SE})$ 
for the scaling 
function from Eq.\ (\ref{eq:fnum})  is 
  shown as green line. 
  For small pressures, the energy barrier scales as $p^{-2.8}$,
   which is close to the $p^{-3}$ law from the Pogorelov approximation
  \eqref{FbarrTh}. 
In the collapse we also inserted 
the experimental data from Ref.\ \cite{Marthelot2017} (yellow squares),
which show excellent agreement with our numerical results.
}
\label{fig:gamma-E}
\end{center}
\end{figure*}

In the simulation, we construct a spherical rest shape
and apply a pressure $p<p_c$, which leads to a uniformly 
compressed sphere with radius $R(p)<R_0$. 
In order to map out the energy barrier between the compressed 
spherical state and the final buckled state, we need to stabilize all 
intermediate transition states in the numerical simulation.
This is done by selecting  two points on opposite 
sides of the sphere and 
introducing an additional constraint 
 on the distance $z$ between these two points
 during energy minimization (corresponding to an indentation depth 
 $\zeta = R - z/2$).
Constraining and decreasing the distance $z$ between these two points 
allows  us  to control the size of the two dimples that 
are formed on opposite sides of the sphere. 
For sufficiently small dimples, the enthalpy of a single dimple 
is  half of the enthalpy of  two dimples
\cite{Hutchinson2017}, as we numerically verify in 
 Fig.\ \ref{fig:2dimple} in 
Appendix \ref{sec:SE}.
This allows us to obtain the enthalpy landscape 
$F=F(z)$ of a single dimple as a function of $z$  
for a given pressure $p$ (see Fig.\ \ref{fig:Fz}), 
from the maximum of which we can directly 
determine the  energy barrier  $F_\text{B}$.

As has been shown above, 
energies of deformed shells only depend on two parameters
$\gamma$ and $\nu$ characterizing the elasticity and the 
dimensionless  pressure  $p/p_c$. 
This also applies to the energy barrier, which, if measured in units 
of $YR_0^2$, should only depend on these parameters. 
To analyze simulation results we use the critical pressure $p_\text{c,SE}$ 
as obtained from SURFACE EVOLVER simulations rather than the 
theoretical value  $p_c = 4 (Y/R_0)\gamma^{-1/2}$  from Eq.\ (\ref{eq:pcb})
in order to compensate for discretization effects from triangulation 
of the surface, as explained in Appendix \ref{sec:discrete}.
The Pogorelov approximation (\ref{FbarrTh})
suggests that the energy barrier can be written 
as a product
\begin{align}
 \frac {F_\text{B}(p)}{YR_0^2}
   = f_\nu(\nu)\gamma^{-\alpha} f_p\left(\frac{p}{p_c}\right),
\label{eq:Fbarrscaling}
\end{align}
with a function $f_\nu(\nu)$, an exponent $\alpha$, 
 and a scaling function $f_p(x)$ for the pressure-dependence,
which has to be determined from the numerical simulation results. 
Our numerical  results for 
 the energy barrier as
a function of $p/p_\text{c,SE}$ for several values of $\gamma$ and $\nu$ are
shown in Fig.\ \ref{fig:gamma-E}.
Figure \ref{fig:gamma-E}(a) demonstrates that, for $\nu\le 0.5$, 
 the $\nu$-dependence is very 
weak as in the Pogorelov  approximation [where $f_\nu(\nu) =a(\nu)$, cf.\ 
Eq.\ (\ref{FbarrTh})]
  such that 
we simply  choose $f_\nu(\nu) \approx 1$ for $\nu \le 0.5$.
 Figure \ref{fig:gamma-E}(c) shows that data for different 
$\gamma$ [Fig.\ \ref{fig:gamma-E}(b)] almost perfectly collapse 
for  an exponent  
\begin{equation}
  \alpha = 1.52\approx 3/2,
\label{eq:alpha}
\end{equation}
 which is also 
 in agreement with the  Pogorelov approximation. 
As a result, we can read off the scaling function $f_p(x)$ from
Fig.\ \ref{fig:gamma-E}(c), which shows clear deviations from the 
Pogorelov approximation $f_p(x) \sim x^{-3}$ for  larger $x\le 1$. 
For small $x$ we find $f_p(x) \sim x^{-2.8}$ close to the Pogorelov 
approximation. 
A numerical  approximation $f_\text{p,num}(x)$ for the scaling 
function is given 
in Appendix \ref{FBarrNum} in  Eq.\ (\ref{eq:fnum}).
Numerically, we see that the energy barrier vanishes 
upon approaching the critical pressure $p_c$, as opposed 
to  the Pogorelov energy barrier  (\ref{FbarrTh}). 
Our data are fairly well described by 
 $F_\text{B}(p)\propto (1-p/p_c)^2$ for $p\approx p_c$ [or 
$f_\text{p,num}(x) \propto (1-x)^2$ for $x\approx 1$].

Our numerical results are in excellent agreement with experimental 
data from Ref.\ \cite{Marthelot2017}, where the energy barrier 
has been determined from experiments on hemispherical shells 
subject to both compressive pressure and a probing point force
controling  depth of the indentation. A comparison with 
the rescaled 
experimental data from  Ref.\ \cite{Marthelot2017} is also 
shown in Fig.\ \ref{fig:gamma-E}(c).

\begin{figure}
\begin{center}
\includegraphics[width=1\linewidth]{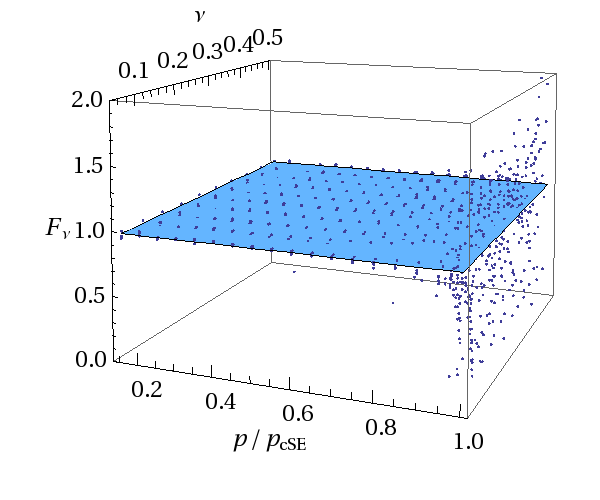}
\caption{
  Energy barrier rescaled according to  Eq.\ (\ref{eq:ratio}) 
  as a function of $\nu$ and $p/p_\text{c,SE}$ for $\nu\le 0.5$ (dots).
  Additionally, a linear fit $f_\nu(\nu) = {\rm const} + 0.086\nu$ 
   is shown
  (plane). The values for very large pressures scatter
    because of the numerical errors in the determination of the 
  critical pressure $p_\text{c,SE}$ and 
   because the determination of  energy barriers 
   for large pressures  or small dimples becomes less exact, as
   small changes to the sphere can cause
  buckling in this regime. Therefore,
    only values $p/p_\text{c,SE}<0.8$ are used for fitting
   $f_\nu(\nu)$.
     }
\label{fig:nu-iso}
\end{center}
\end{figure}

In order to determine the $\nu$ dependence more exactly, i.e., 
to obtain a more accurate approximation for 
 the function $f_\nu(\nu)\approx 1$ in Eq.\ (\ref{eq:Fbarrscaling}), 
we perform simulations
 for different values of $\gamma$ and
different values of $\nu$, and isolate $f_\nu(\nu)$ by 
plotting the rescaled energy barrier
\begin{align}
  \frac {F_\text{B}(p) \gamma^{1.52}} 
  {YR_0^2f_\text{p,num}(p/p_\text{c,SE})} = f_\nu(\nu)
\label{eq:ratio}
\end{align}
in Fig.\ \ref{fig:nu-iso}, where  we use the scaling 
function $f_\text{p,num}(x)$ from Fig.\ \ref{fig:gamma-E}(c). 
For $\nu \le 0.5$, a linear fit 
$f_\nu(\nu) = 0.98 + 0.086\nu$ describes the data
[see also Eq.\ (\ref{eq:fnu})].

Using the numerically determined functions $f_\text{p,num}(x)$ and 
$f_\nu(\nu)$, we obtain an accurate description of the numerical 
results, which is summarized in Appendix \ref{FBarrNum}
in Eq.\ (\ref{eq:FBarrnum}). 
A simple approximation formula, which is accurate to within $20\%$,
is obtained using $f_\nu(\nu)\approx 1$, $\alpha=3/2$,  and
\begin{align}
 \frac {F_\text{B}(p)}{YR_0^2} 
   &\simeq \gamma^{-3/2} f_\text{p,app}\left(\frac{p}{p_c}\right)
\nonumber\\
 f_\text{p,app}(x) &=
    1.44\,(1-x)^2 (x^{-3} +34.1\,x^{-1}).
\label{eq:fapp}
\end{align}
The approximative scaling function $f_\text{p,app}(x)$ is 
motivated by the Pogorelov result 
$F_\text{B}(p) \propto p^{-3}$ for small  $p\ll p_c$ and 
our numerical result $F_\text{B}(p) \propto (1-p/p_c)^{2}$
for $p\approx p_c$. 
The $x^{-1}$-term represents a $p^{-1}$ correction to the 
Pogorelov result.

\subsection{Critical pressure from thermal activation}

Our results for the energy barrier can be used to estimate 
a time scale on which buckling will occur by thermal activation.
According to Kramers theory, i.e., assuming effectively
overdamped stochastic dynamics for a reaction coordinate characterizing 
the size of the dimple such as $z$ or $\Delta V$, 
 this time scale  is
\begin{equation}
   t_{B} = \tau e^{F_\text{B}(p)/k_B T},
\end{equation}
where $\tau_0$ is a microscopic time scale characterizing 
the dynamics of the reaction coordinate and where the
 Arrhenius factor $\exp(F_\text{B}/k_B T)$ dominates.
For  a given  experimentally accessible time scale $t_\text{exp}$
buckling occurs if 
\begin{equation}
  F_\text{B}(p) < k_BT \ln(t_\text{exp}/\tau) \sim k_BT,
\end{equation}
which is essentially the criterion that energy barriers  
$F_\text{B}(p) < k_BT$
can be overcome by thermal activation quasi-spontaneously
(as long as they are 
not several orders of magnitude between time scales 
$t_\text{exp}$ and $\tau$). The criterion 
$F_\text{B}[p_{c}(T)] = k_BT$ then 
determines an effective temperature-dependent 
buckling pressure $p_{c}(T)$ for buckling by thermal activation.

First, we use this criterion with our $T=0$ results for the buckling 
energy barrier from the previous section. Employing the 
Pogorelov approximation  \eqref{FbarrTh} we obtain 
\begin{equation}
  \frac{p_{c}(T)}{p_c} =   a^{1/3}(\nu) \left(\frac{k_BT}{\kappa}\gamma^{1/2}
  \right)^{-1/3}  =   a^{1/3}(\nu) \bar{T}^{-1/3}
\label{eq:bcT1}
\end{equation} 
for $\bar{T} \gg 1$ such that $p_{c}(T) \ll p_c$. 
This result is only valid where the Pogorelov approximation is 
valid, i.e., for $0.8 >p_{c}(T)/p_c  \gg \gamma^{-1/4}$ as discussed
above, which means $0.8^{-3}  < \bar{T}  \ll \gamma^{3/4}$ in
Eq.\ (\ref{eq:bcT1}). 

\begin{figure}
 \includegraphics[width=1\linewidth]{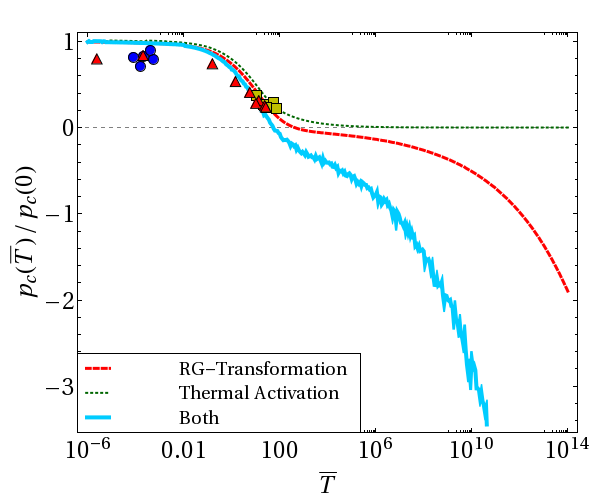}
\caption{
  Critical buckling pressure $p_c(T)$  as a function of the dimensionless
  temperature $\bar{T}= {k_BT\sqrt{\gamma}}/{\kappa} $ 
   according to the  buckling criterion $D_R(q^*)=0$  which includes 
renormalization effects only 
(dark red dashed line, see also red solid line in 
Fig.\ \ref{fig:pcs}), 
$p_{c}(T)$ according to the criterion $F_\text{B}[p_{c}(T)] = k_BT$
or Eq.\ (\ref{eq:bcT2}) 
for thermal activation over the buckling energy barrier
 (green dotted line, for $\nu=1/3$), and 
$p_{c}(T)$ according to the 
generalized criterion ${F_\text{B,R}(q^*)}/{k_B T} = 1$ or Eq.\ 
(\ref{eq:bcT3}) which combines renormalization 
effects and thermal activation (light blue solid line, for $\nu=1/3$). 
Inclusion of thermal activation leads to a decrease of the 
critical buckling pressure  for $\bar{T}>1$. 
 As  in Fig.\ \ref{fig:pcs}, data points show numerical 
results from Ref.\ \cite{Paulose2012}. 
}
\label{fig:Fbarr}
\end{figure}

Using the numerical results for the energy barrier in the scaling form 
${F_\text{B}(p)}/{YR_0^2} 
  \approx f_\nu(\nu)\gamma^{-3/2} f_\text{p,num}(x)\left({p}/{p_c}\right)$
 with the scaling 
function $f_\text{p,num}(x)$  given by the collapse in 
Fig.\ \ref{fig:gamma-E}(c), we obtain the buckling criterion
\begin{equation}
   f_\text{p,num}\left(\frac{p_{c}(T)}{p_c}\right)  
    = \frac{1}{f_\nu(\nu)} \frac{k_BT}{\kappa}\gamma^{1/2}
    =  \frac{\bar{T}}{f_\nu(\nu)}
\label{eq:bcT2}
\end{equation} 
for buckling by thermal activation.
The resulting critical buckling pressure  $p_c(T)$ is shown in Fig.\ 
\ref{fig:Fbarr} (dark red line) and lies slightly above the 
critical buckling pressure  from parameter renormalization. 
In the absence of thermal fluctuations, 
 the right-hand side vanishes and Eq.\ (\ref{eq:bcT2}) reduces 
to the correct result 
$p_{c}(0)/p_c=1$ because this is the pressure where also 
the barrier function $f_\text{p,num}\left({p_{c}(0)}/{p_c}\right)$ 
on the left-hand side vanishes. 
With $f_\nu(\nu)\approx 1$
and   $f_\text{p,num}(x) \approx 
    1.44\,(1-x)^2$ [see Eq.\ (\ref{eq:fapp})],
 we can also infer the crossover to the $T=0$ result:
 $p_{c}(T)/p_c \approx 1- (\bar{T}/1.44)^{1/2}$.
Interestingly, also the criterion (\ref{eq:bcT2}) for buckling 
by thermal activation only depends on the  
dimensionless temperature $\bar{T}$ as also observed for 
the influence of  parameter renormalization on 
the buckling instability; see Fig.\ \ref{fig:pcs}.

So far, we did not take  the combined effects of 
 {\it both} thermal activation {\it and} renormalization  of the 
parameters $\kappa$, $Y$, and $p$ by anharmonicities 
into account  in the buckling criteria.
This can be done by using properly renormalized parameters 
$\kappa_R$, $Y_R$, and $p_R$ in the 
 barrier crossing  criterion 
$F_\text{B}(p) = k_BT$, which leads to  
\begin{equation}
   f_\text{p,num}\left(\frac{p_R(q^*)}{p_{c,R}(q^*)}\right)  
    =   \frac{\bar{T}}{f_\nu(\nu)},
\label{eq:bcT3}
\end{equation} 
where $p_R(q^*)$ is the renormalized pressure  and $p_{c,R}(q^*) = 
{4\sqrt{\kappa_R(q^*) Y_R(q^*)}}/{R_0^2}$ is 
the renormalized critical buckling pressure.
The renormalization of $\nu$ need not be considered, because of
the negligible effect of $\nu$ on the energy barrier, and because $\nu$
remains of order unity during the renormalization
\cite{Aranovitz_Lubensky}.
The critical buckling pressure $p_{c}(T)$ in the presence 
of thermal activation is the smallest initial 
pressure for which criterion (\ref{eq:bcT3}) can be fulfilled.

 The question remains to be answered 
how to choose the length scale $l^*= {\pi}/q^*$   in Eq.\ (\ref{eq:bcT3})
up to which barrier parameters are renormalized in the 
modified criterion   ${F_\text{B,R}(q^*)}/{k_B T} = 1$.
The transition state at the energy maximum  is a sphere
with  an energetically unfavorable  ``flattened'' dimple 
(see also Fig.\ \ref{fig:Fz})
and the elastic parameters characterizing this transition state 
dimple can only be renormalized by fluctuations on length scales smaller
than the size of the dimple.  
At finite temperatures, the minimal transition state 
dimple size is set by the unstable wave number $q^*$ 
for buckling, which we
determined by our above  buckling criterion $D_R(q^*)=0$ or the 
approximative buckling criterion  $p_R(q^*)=p_{c,R}(q^*)$  
from Ref.\ \cite{Kosmrlj2017} from parameter renormalization.
At $T=0$, the buckling  length scale
 $l^*$ approaches  the  elastic length scale 
$l_\text{el}\sim R_0\gamma^{-1/4}$  for the  unstable wave length at the 
$T=0$ buckling instability (see Fig.\ \ref{fig:lengthscale}).
At $p=p_c$ and $T=0$, the elastic length scale $l_\text{el}$ also sets 
the size of the transition state dimple.
 For $p_{c} >p>p_{c1}$, the $T=0$ 
transition state dimple size can be obtained from 
 the enthalpy landscape $F(p,\Delta V_b,\Delta V)$, for which we 
showed above that the barrier state is assumed for an indentation volume 
$(\Delta V-\Delta V_b)/V_0  \sim \gamma^{-1}(p/p_c)^{-4}$. Assuming a 
spherical cap shaped dimple, this corresponds to 
a transition state dimple size  $\sim R_0 \gamma^{-1/4}p_c/p$, which 
increases with decreasing pressure.
It reaches its maximally possible size $R_0$ at the Maxwell
pressure $p_{c1}$ above which  buckling becomes energetically possible
and reduces to $l_\text{el}\sim R_0 \gamma^{-1/4}$ at $p=p_c$. 
Therefore, also at finite temperatures, we  expect the 
transition state dimple size to be 
somewhat larger than $l^*$ for $p<p_c(T)$. 
At $T=0$, the length scale $l_\text{el}$ remains the length scale of the 
width of the ridge of the Pogorelov dimple \cite{Knoche2014,Knoche2014a}
also for $p<p_c$. 
In analogy with the $T=0$ case, we  expect that the length scale 
$l^*$ determined by our above  buckling criterion $D_R(q^*)=0$
corresponds to the width of the Pogorelov dimple ridge 
in the presence of thermal fluctuations rather 
than the size of the dimple.
Thus, in choosing 
$q^*$  from the 
buckling criterion $D_R(q^*)=0$ in Eq.\ (\ref{eq:bcT3}), 
we use parameters, which are renormalized 
only up to the length scale of the  ridge of the dimple.

In using renormalized parameters in 
the buckling criterion (\ref{eq:bcT3}), which includes 
thermal activation, we also  assumed
 that  the same RG equations (\ref{BetaFunctions}) that are derived 
for fluctuation around a  spherical background state can still be applied 
to describe fluctuations around the  transition 
state  which already contains the ``flattened'' dimple.
The RG equations will most likely take a different form 
on length scales exceeding the Pogorelov ridge scale $l^*$
 because stretching and bending
strains with this wavelength are large in a dimple configuration.
Stretching and bending modes with smaller wave lengths 
should remain small and unaffected by  the dimple
such that they are well described by the  RG equations (\ref{BetaFunctions})
for a  spherical background.

If the length scale $q^*$ is chosen from 
the  approximative buckling criterion  $p_R(q^*)/p_{c,R}(q^*)=1$  
from Ref.\ \cite{Kosmrlj2017},
Eq.\ (\ref{eq:bcT3}) can be written as $p_R(q^*)/p_{c,R}(q^*)  
    =  f_\text{p,num}^{-1}[\bar{T}/f_\nu(\nu)]$ (with $f_\text{p,num}^{-1}$ 
as inverse function of  $f_\text{p,num}$)
and viewed as direct generalization 
of this criterion in the presence of thermal activation. 
The combined effect of thermal activation and parameter renormalization 
then 
leads to a further reduction of the critical buckling pressure, in particular, 
for temperatures $\bar{T}>1$. 
This is what we expect in general:
If both buckling mechanisms are considered,
 the critical buckling pressure  should be further lowered 
below our above results in  Fig.\ \ref{fig:pcs}
 from parameter renormalization only.

If the length scale $q^*$ is chosen from our buckling 
criterion $D_R(q^*)=0$, we obtain the generalized 
critical pressures $p_{c}(T)$ in Fig.\ \ref{fig:Fbarr} (blue solid line).
Again, 
the combined effect of thermal activation and parameter renormalization 
leads to a further reduction of the critical buckling pressure
for temperatures $\bar{T}>1$. 
 The available numerical Monte-Carlo 
simulation  results from Ref.\ \cite{Paulose2012}, which are also shown 
in Fig.\ \ref{fig:Fbarr}, do not lie in the temperature range $\bar{T}\gg 1$,
where the additional  reduction becomes most pronounced
demonstrating the need for further simulations in this temperature
range.

\subsection{Buckling shapes}
\label{sec:shapes}

\begin{figure}
 \includegraphics[width=1\linewidth]{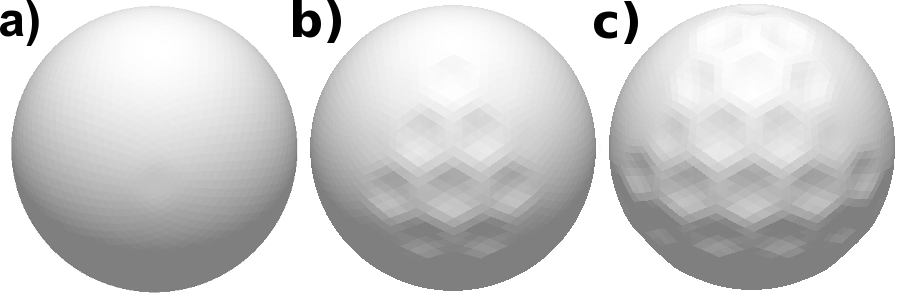}
\caption{
 Postbuckling shapes 
obtained from numerical energy minimization with a constrained 
variance  $\Delta R^2$  and 
   unconstrained $z$
for (a) $\Delta R^2/R_0^2 = 1.3\times 10^{-7}$,
 (b) $\Delta R^2/R_0^2 = 1.5\times 10^{-5}$, and 
(c)  $\Delta R^2/R_0^2 = 4.7\times 10^{-5}$  for $p/p_c=0.85$.
}
\label{fig:Postbuckling}
\end{figure}

So far only spheres with one or two dimples have been considered. However, as
the mode that becomes unstable for $p\ge p_c$ produces multiple dimples,
and, in fact, the initial buckling shapes are a combination of critical modes
\cite{Hutchinson1967, Koiter1969}, 
as shown in Fig.\ \ref{fig:Postbuckling},
such shapes must also be taken into account.
The question remains to what  extent configurations with multiple 
dimples represent also relevant intermediate states 
for the energy barrier  for $p<p_c$, i.e., 
whether they represent the optimal transition states between 
 the metastable compressed spherical state and the buckled state.
The final buckled state will have only a single dimple. 
 
\begin{figure*}
\begin{center}
\includegraphics[width=0.99\linewidth]{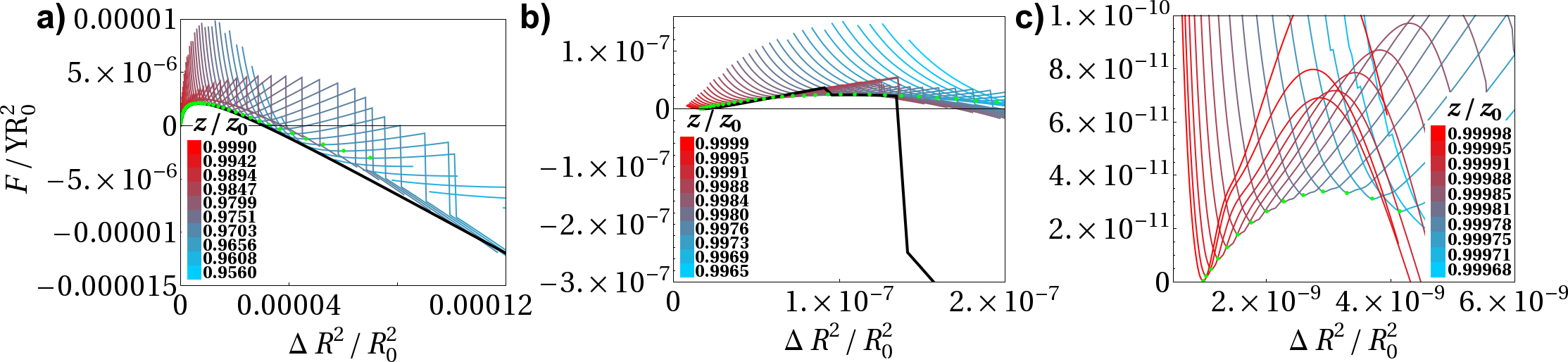}
\caption{
Minimal enthalpy as a function of $\Delta R^2$ obtained by the two
  outlined approaches, i.e., for constrained $z$ (colored lines)
   or unconstrained $z$ (thick black lines)
     for $\gamma = 100000$, $\nu=0.3$, and $p/p_c =
  0.3$ (a),  $p/p_c = 0.85$ (b),  and 
   $p/p_\text{c,SE} \approx 0.998$ (c). 
  Lines of different colors
  correspond to different values of $z$. Green dots show local energy minima
  for different values of $z/z_0$ with $z_0$ being the initial value of $z$,
  before deformation.
(a) The colored lines of  the $z$-constrained 
   enthalpy touch the black line for unconstrained $z$ (i.e., the 
  black line is the envelope of the colored lines) indicating that  the 
  optimal buckled configuration  consists of a single dimple.
(b) Jumps in the colored and black curves
are caused by large inward moves of a vertex in the center of the 
indentation when the indentation switches 
from its initial  concave  into its final  convex  shape. 
These jumps occur behind the initial energy barrier. 
The black line  shows small 
deviations from  the envelope 
of the colored lines for small  $\Delta R^2$ because
the location of a single dimple is determined by the 
$z$-constraint (colored lines), whereas the 
initial formation and location of dimples is 
strongly influenced by the presence of disclinations 
 for unconstrained $z$ (black line). 
(c)  Close to the critical pressure the influence of the 
disclinations is even stronger (therefore, data  for unconstrained
$z$ is not shown).  
}
\label{fig:Shapes}
\end{center}
\end{figure*}

In order to compare enthalpies of states with single dimples 
and states with multiple dimples, we use two types of constraints. 
The first type is a constraint on the 
variance  $\Delta R^2$ of the distance of every vertex on the sphere to
its center (the  center of the sphere is considered to be 
 the average of the coordinates of all vertices).
  The variance  $\Delta R^2$ of a spherical state 
before buckling is zero, whereas $\Delta R^2>0$
 after buckling, both for a single dimple and multiple 
dimples (see Fig.\ \ref{fig:Postbuckling}).
By constraining the variance, we  stabilize 
 buckling shapes both with single and multiple dimples. 
The second type of constraint  is similar
to the constraint    on the distance $z$ between two opposite 
vertices on the sphere, which we used in Sec.\ \ref{Simulation}.
Now, we fix, however, the position of one vertex and constrain the other 
vertex to a  distance $z$.
For a constrained $z<2R_0$ a single dimple of 
depth $\zeta = 2R_0-z$ will be formed.
We  apply both constraints to the buckled shapes and compare 
the enthalpy $F$ for different target values  for $\Delta R^2$.
If the minimal enthalpy for a given target value of $\Delta R^2$ 
without an additional $z$ constraint 
 can also be reached in the presence of a $z$ constraint, we can conclude
that this enthalpy minimum represents a configuration with a single 
dimple. If, on the other hand, the 
minimal enthalpy for a given target value of $\Delta R^2$ 
without an additional $z$ constraint 
 cannot  be reached in the presence of a $z$-constraint, 
we conclude that  this enthalpy minimum is a configuration 
with multiple dimples.

 The results for the enthalpy minima as a function of  $\Delta R^2$ 
 are shown in Fig.\  \ref{fig:Shapes}. The black line is the 
minimal enthalpy without additional $z$ constraint, and the colored lines 
are enthalpy minima in the presence of both constraints with 
color-coded $z$-value. 
The black line or the envelops of the colored lines 
 exhibit a maximum which is the enthalpy barrier 
for buckling. 
If the colored lines remain
   above the black line  for unconstrained $z$, the 
  optimal buckled configuration  consists of an array of multiple 
 dimples. If  the colored lines of  the $z$ constrained 
   enthalpy touch the black line for unconstrained $z$ (i.e., the 
  black line is the envelope of the colored lines), we conclude that 
 the 
  optimal buckled configuration  consists of a single dimple.
 The results in Fig.\ \ref{fig:Shapes} show that 
the  latter is the case.

\begin{figure}
 \includegraphics[width=1\linewidth]{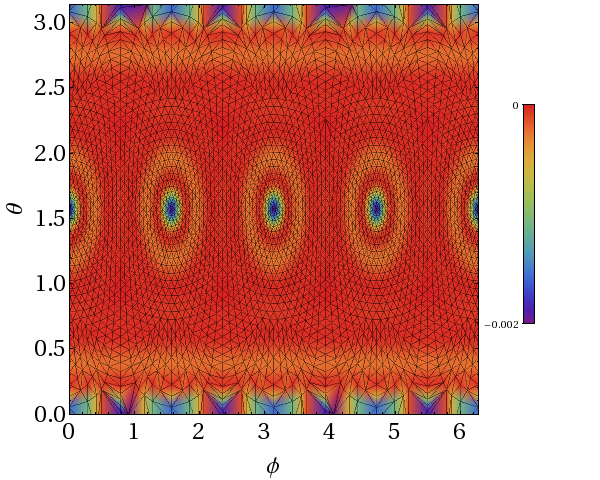}
\caption{
 $R(\phi, \theta)-\bar R$, in spherical coordinates, 
 where $R(\phi, \theta)$ is the
  distance to the center of the sphere, and $\bar R$ is its average, 
  for an  additional $z$ constraint  and directly before
  the first jump in Fig.\ \ref{fig:Shapes}(b)  for $p/p_c = 0.85$.
  The variance  $\Delta R^2$
  is mostly caused by six fourfold disclinations
 (four at $\theta\sim \pi/2$
and one at $\theta=0,\pi$), such that the triangulation determines 
   the formation and location of six dimples instead of a single dimple. 
As a result,  the black line in   Fig.\ \ref{fig:Shapes}(b) 
does not envelope the colored lines in   Fig.\ \ref{fig:Shapes}(b)
for  small $\Delta R^2$.
 }
\label{fig:ShapesDisc}
\end{figure}

There are small deviations, where the envelope of the colored lines 
does not coincide with the black line at larger pressures $p/p_c = 0.85$
in Fig.\ \ref{fig:Shapes}(b). The deviations at small 
 $\Delta R^2$ are caused by the effect of unavoidable disclinations in the 
triangulation of the spherical surface. 
We use a triangulation that contains six fourfold disclinations
(each carrying a topological charge of 2, producing an overall charge of 12);
 see Appendix \ref{sec:SE}.
Indentations interact 
with these disclinations and are initially formed at disclinations. 
Therefore, we observe six very shallow indentations which form 
at disclinations 
in the absence of a $z$ constraint along the black line, as also 
illustrated in Fig.\ \ref{fig:ShapesDisc}. With increasing 
 $\Delta R^2$, the system switches to a single indentation.  The interaction 
of indentations with disclinations is stronger for large pressure.

\begin{figure}
 \includegraphics[width=1\linewidth]{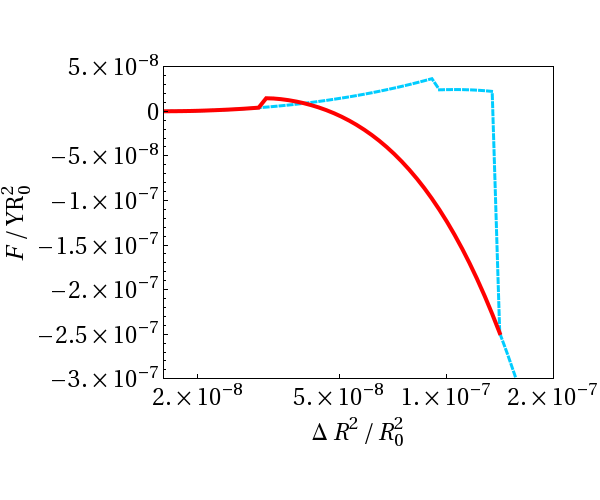}
\caption{Shape hysteresis and 
  minimal enthalpy as a function of $\Delta R^2$. 
  The light blue dashed line
    shows the minimum enthalpy for {\it increasing} $\Delta R^2$,
    starting from the spherical shape, and the dark red solid line
    shows the minimum
    enthalpy found for {\it decreasing} $\Delta R^2$, starting with a
    concavely indented shape. The reverse red path with 
  concavely indented shapes exhibits a lower global
    enthalpy maximum than the blue forward path and becomes unstable 
  for small $\Delta
    R^2$, where shapes become  convex again.
}
\label{fig:Lower_FB}
\end{figure}

\begin{figure}
 \includegraphics[width=1\linewidth]{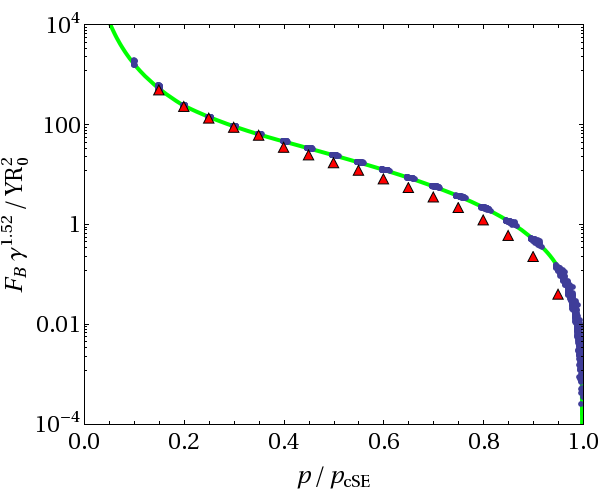}
\caption{Collapse of the energy barrier obtained by applying an
  additional   $z$ constraint and for increasing $\Delta R^2$ (dots), 
  numerical approximation of these values (line), and
    the slightly lower 
   energy barrier (red triangles) obtained by reducing  $\Delta R^2$ 
 starting from a concave, indented shape.
}
\label{fig:Lower_FB2}
\end{figure}

There are also downward energy jumps for  $p/p_c = 0.85$ 
in Fig.\ \ref{fig:Shapes}(b). 
These jumps are caused by large inward moves of a vertex in the
center of the indentation when the indentation switches from its initial
unstable convex shape into its final concave shape.  
For pressures $p/p_c \approx 0.5$,
the constraint on the variance $\Delta R^2$ is not sufficient 
to suppress such sudden switches, eventually also as an effect of 
the existence of several small indentations because of the presence 
of disclinations, which affect $\Delta R^2$.
Associated with this sudden switch is  a hysteretic behavior 
if $\Delta R^2$ is reduced again starting from a concavely indented
post-jump state. Then the capsule returns into its initial spherical shape 
along a different path of configurations with different energies 
as shown in Fig.\ \ref{fig:Lower_FB} also for $p/p_c = 0.85$. 
The reverse deformation path exhibits a slightly smaller energy barrier. 
 Despite the lower energy barrier, this deformation path has a
steeper initial rise in energy when starting from the spherical shape, and is
therefore not taken when the capsule is gradually indented. The energy
barrier measured on the reverse path 
is only slightly lower, as the comparison  in
Fig.\ \ref{fig:Lower_FB2} shows, 
and will therefore have  no significant influence on
any of the previously discussed results.

For pressures that approach $p_c$, the paths with decreasing energy
move closer to the energy maximum, as shown in Fig.\ \ref{fig:Shapes}(c).

Only if we further follow the system to much larger values of $\Delta R^2$ 
{\it behind}  the energy barrier, multiple dimples are formed  as shown in 
Fig.\ \ref{fig:Postbuckling}. These dimples start to form around the 
initial single dimple and spread over the entire sphere.

We conclude that single dimples appear to be stable {\it before}
the energy barrier, i.e.,  as long as dimple formation still increases 
the enthalpy, whereas multiple dimples spread across the sphere 
{\it behind} the energy barrier where further dimple formation 
lowers the enthalpy again. 
Then the path with decreasing energy, along which multiple dimples 
can be formed,  appears almost directly behind
the energy maximum. This
 is consistent with the known behavior at the critical buckling 
pressure  $p=p_c$. Then 
the energy maximum moves to $\Delta R^2 = 0$ and 
multiple dimples can immediately be formed \cite{Hutchinson1967}.

\section{Conclusions}

Thermal fluctuations lead to a depression of the critical buckling 
pressure $p_c$ by two mechanisms: (i) parameter renormalization 
because of 
anharmonicities in normal displacement modes, which are mainly caused 
by the background curvature $1/R_0$ of the shell 
and (ii) thermal activation over the buckling barrier. 
In Ref.\ \cite{Kosmrlj2017},  the effects of parameter renormalization 
have already been considered; here we extended this approach to also 
take thermal activation into account.

The anharmonic coupling between different normal displacement modes 
leads to scale-dependent, renormalized  elastic moduli 
$\kappa_R(q)$ and $Y_R(q)$, and also a scale-dependent pressure 
$p_R(q)$, which can be obtained from a RG transformation that has been 
derived in Ref.\ \cite{Kosmrlj2017}. 
Buckling sets in if the two-point vertex function 
$D_R(q)$, which characterizes the curvature of the effective 
harmonic free enthalpy for the normal 
 deformation mode $\tilde f(q)$, is renormalized to zero for a 
certain wave number $q=q^*$: $D_R(q^*)=0$. 
The mode $\tilde f(q^*)$ then becomes
unstable  and initiates the  buckling transition. 
This buckling criterion is slightly different from the buckling 
criterion $p_R(q^*)/p_{c,R}(q^*)=1$, which is based
 on the  renormalized ratio of pressure to critical 
buckling pressure that has been used  in Ref.\ \cite{Kosmrlj2017}.
The resulting temperature-dependent critical buckling pressures 
are summarized in Fig.\ \ref{fig:pcs}. The criterion $D_R(q^*)=0$
gives a slightly higher value than obtained in  Ref.\ \cite{Kosmrlj2017}.

The main conclusions regarding renormalization effects remain unchanged
from Ref.\ \cite{Kosmrlj2017}.
 Renormalization  effects 
should be most pronounced for large dimensionless temperatures 
$\bar{T}=k_BT\sqrt{\gamma}/\kappa \sim (k_BT/\kappa) \sqrt{YR_0^2/\kappa}$
[see Eq.\ (\ref{eq:barT})].
For shells made from thin isotropically elastic materials 
this parameter becomes $\bar{T}\sim k_BT R_0/Eh^4$, which  means 
that shells  with large radius $R_0$
are more susceptible to thermal fluctuations.
The relevance of thermal fluctuations 
 depends most sensitively ($\bar{T}\propto h^4$) on shell thickness,
and thin shells are most susceptible to thermal fluctuations.

For a typical soft material with $E \sim 0.1\,{\rm GPa}$, from which 
a microcapsule of size $R_0\sim 10\,{\rm \mu m}$ 
is synthesized, the shells should 
be ultrathin with $h \sim 1\,{\rm nm}$ to reach $\bar{T} \sim 400$. 
Also many biological capsules, such as red blood cells ($\bar{T} \sim 2-40$) 
or Gram-negative bacteria ($\bar{T} \sim 8$), can exhibit 
fairly large $\bar{T}$.
Moreover, $p_c(T)$ can, in principle, become negative also with the 
modified buckling criterion $D_R(q^*)=0$, as 
Fig.\ \ref{fig:pcs} clearly shows. Therefore, a  spontaneous buckling
{\it without} external pressure but
 only caused by the fluctuation-generated
pressure is possible as discussed in Ref.\ \cite{Kosmrlj2017}.

We extended the Pogorelov approximation and 
used numerical energy minimization with the SURFACE EVOLVER to 
quantify the energy barrier $F_\text{B}(p)$ 
for a single axisymmetric dimple.
We found a scaling behavior ${F_\text{B}(p)}
   = YR_0^2 f_\nu(\nu)\gamma^{-\alpha} f_p\left({p}/{p_c}\right)$ [see 
Eq.\ (\ref{eq:Fbarrscaling})]
with an exponent $\alpha \simeq 3/2$ and a weak $\nu$ dependence 
$f_\nu(\nu) \approx 1$, in agreement with the 
Pogorelov approximation, i.e., 
$F_\text{B}(p) \sim (Eh^4/R_0) f_p\left({p}/{p_c}\right)$.
  We quantified the scaling function $f_p(x)$ 
numerically in Fig.\ \ref{fig:gamma-E} in excellent 
agreement with existing experimental data from Ref.\ \cite{Marthelot2017}.
A simple approximation is given by 
Eq.\ (\ref{eq:fapp}) with a 
 pressure scaling function  $f_p(x) \sim (1-x)^2x^{-3}$, which is 
 in agreement with the Pogorelov approximation $F_\text{B} \propto p^{-3}$
for $x\ll 1$ and 
predicts that the barrier vanishes $F_\text{B} \propto (1-p/p_c)^2$ 
close to the critical pressure. The exponent 2 is a numerical result 
and has to be corroborated by theoretical arguments in future research. 

Considering a capsule or shell  at half critical pressure $p= p_c/2$ (with 
$f_p(0.5) \sim 50$),  
we find $F_\text{B}(p_c/2) \sim k_BT  \sim  50 /\bar{T}$ and  the barrier 
height measured in thermal energy units is also governed by the
dimensionless temperature $\bar{T}$. 
Instead of using the thermal energy scale $k_BT$, we can also 
use a mechanical energy scale and state that
the characteristic size of the buckling barrier, 
 $F_\text{B}(p_c/2) \sim YR_0^2 \gamma^{-3/2}\sim \kappa
\gamma^{-1/2} \sim Eh^4/R_0$, 
is reduced over  the elastic compression 
energy of the spherical shell just before buckling, 
$U_{\rm sph} \sim \kappa\sim Eh^3$, by a factor $\gamma^{-1/2}\sim h/R_0$.

For $p<p_c$, this energy barrier can be crossed by thermal 
activation on accessible time scales if 
$F_\text{B}<k_BT$, which serves as criterion 
for thermal buckling via thermal activation. 
Thermal activation leads to a similar depression of $p_c(T)/p_c<1$ 
as parameter renormalization. 
Both approaches give a critical pressure $p_c(T)/p_c$ 
which only depends on the dimensionless temperature 
$\bar{T}$. 
Finally, we combine parameter renormalization effects and thermal activation 
by using properly renormalized  elastic parameters for the 
energy barrier. This leads to our final results in Fig.\ \ref{fig:Fbarr},
which shows that thermal activation leads to a significant further 
decrease in $p_c(T)$ and cannot be neglected:
For the relative difference  $\Delta p_c(T)$  between 
the critical pressures  from renormalization only
and from thermal activation plus renormalization 
as compared to  the zero temperature critical pressure $p_c(0)$
we find  $|\Delta p_c(T)|/p_c(0) = 7,~11,~17,~27 \%$ 
for dimensionless temperatures $\bar{T} =1,~10,~100,~1000$.

Finally, we addressed the question to what extent buckling shapes 
with multiple dimples, which are known to govern the classical 
buckling instability  at $p=p_c$, also play a role for the energy barrier 
for $p<p_c$ by constrained numerical energy minimization. 
In order to allow for the formation of multiple dimples we employed
a constraint on the variance  $\Delta R^2$ of the distance of vertices 
on the shell to the center of the shell. We compared with 
energy minimization where we constrain the distance of one vertex 
to the center, which only leads to formation of a single dimple. 
Our numerical results show 
that single dimples are stable {\it before}
the energy barrier, whereas multiple dimples spread across the sphere 
{\it behind} the energy barrier when further dimple formation 
lowers the enthalpy again. 
Therefore, thermal activation at $p<p_c$ is governed by 
formation of a single indentation.

\begin{acknowledgments}
This research was supported in part by the National Science
Foundation under Grant No. NSF PHY11-25915.
We would also like to acknowledge useful discussions with 
Andrej Ko\u{s}mrlj. 
\end{acknowledgments}

\appendix

\renewcommand{\theequation}{A\arabic{equation}}

\section{Renormalization group calculation}
\label{diagramsApp}

Following  Ko\u{s}mrlj and Nelson \cite{Kosmrlj2017}, we re-derive 
the RG equations governing the scale dependence 
of $\kappa_R(q)$, $p_R(q)$, and $Y_R(q)$.

For a momentum shell RG procedure, we Fourier transform  the effective enthalpy 
\eqref{eq:F}, which results in $F_\text{eff} = F_0 + F_\text{int}$
with a quadratic part 
\begin{align}
  \frac {F_0} A &= \sum_{\bf q} \frac{1}{2}
  \left[\kappa q^4 - \frac {pR_0q^2} 2 
    + \frac Y {R_0^2}\right] \tilde f({\bf q}) \tilde f(-{\bf q})
\label{eq:F0}
\end{align}
and cubic and quartic interactions 
\begin{align}
\frac {F_\text{int}} A &= 
   \sum_{\substack{ {\bf q}_1 + {\bf q}_2 = {\bf q} \ne 0 \\ {\bf q}_3+{\bf
         q}_4 = -{\bf q} \ne 0}}  \frac{Y}{8}
  [q_{1i} P_{ij}^T({\bf q}) q_{2j}][q_{3i}P_{ij}^T ({\bf q}) q_{4j}] \nonumber\\
&~~~~~~~~~~~~~~~~~~\times \tilde f({\bf q}_1)\tilde f({\bf q}_2)
      \tilde f({\bf q}_3)\tilde f({\bf q}_4) \nonumber\\
&~~ + \sum_{\substack{ {\bf q}_1\ne 0 \\ {\bf q}_2 + {\bf q}_3 = -{\bf q}_1}} \frac{Y}{2R_0}
    [q_{2i}P_{ij}^T ({\bf q}_1) q_{3j}] \tilde f({\bf q}_1) \tilde f({\bf q}_2) 
       \tilde f({\bf q}_3),
\label{eq:Fint}
\end{align}
with $P_{ij}^T({\bf q}) =  \delta_{ij} - q_iq_j/q^2$.
The normal  displacements $\tilde f$ are separated into slow modes
 $\tilde f_<({\bf x}) = \sum_{\abs {\bf q} < \Lambda/b} e^{i{\bf q}\cdot{\bf x}}
   \tilde f ({\bf q})$ and fast modes 
 $\tilde f_>({\bf x}) = \sum_{\abs {\bf q} > \Lambda/b} e^{i{\bf q}\cdot{\bf x}} 
   \tilde f ({\bf q})$ containing modes with wave vectors smaller and larger 
than $\Lambda/b$, respectively.

 Integrating out fast  normal displacement modes
$\tilde f({\bf k})$
in the momentum shell $\Lambda/b < |{\bf k}| < \Lambda$  results in
an effective enthalpy 
$F'_\text{eff}[\tilde f_<]= -k_B T 
  \ln \left( \int \mathcal D [\tilde f_>({\bf x})] e^{-F_\text{eff}/k_B
    T}\right)$,
which only depends  on slow normal displacement modes 
with wave vectors  $|{\bf q}| < \Lambda/b$,
\begin{align}
 F'_\text{eff}[\{{\bf q}\}] &= 
   -k_B T \ln \left( \int \mathcal D [\tilde f({\bf k})] 
   e^{-(F_0[\{{\bf q},{\bf k}\}] + F_\text{int}[\{{\bf q},{\bf k}\}])/k_B
     T}\right) 
  \nonumber\\
&= F_0[\{{\bf q}\}]
     - k_B T \ln\braket{e^{-F_\text{int}[\{{\bf q},{\bf k}\}]/k_B T}}_{0,k}.
 \label{eq:Feffq}
\end{align}
The average $\braket{...}_{0,k}$ is an average over fast modes with 
the quadratic part $F_0[\{k\}]$.

\begin{figure}
\begin{center}
\includegraphics[width=1\linewidth]{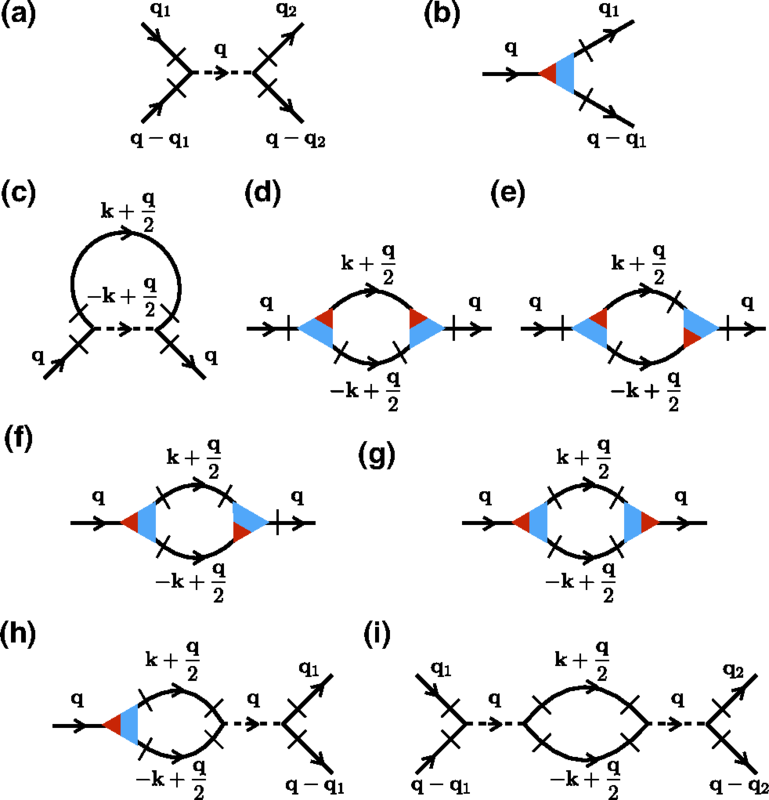}
\caption{
  Feynman diagrams approximating \eqref{F_cumu} to one-loop order.  The
  legs correspond to $\tilde f({\bf q})$, legs with slashes correspond to
  derivatives of $\tilde f({\bf q})$, i.e.,  $q_i\tilde f ({\bf q})$.
  Connected legs
  represent the propagator $G({\bf q})$. The blue parts of the three-point
  vertices connect to the derivatives, whereas the red parts connect to legs
  without slashes. Three-point vertices carry the factor $Y/R_0$, and
  four-point vertices carry the factor $Y$. Image taken from 
  Ref.\  \cite{Kosmrlj2017}.
 }
\label{fig:Feynman}
\end{center}
\end{figure}

The logarithm can be expanded into cumulants, denoted by the superscript $(c)$,
\begin{align}
 F'_\text{eff}[\{{\bf q}\}] &= F_0[\{{\bf q}\}] 
\nonumber \\ 
  &~~~
+ \sum_n \frac{(-1)^{n-1}}{n!(k_B T)^{n-1}} 
   \braket{(F_\text{int}[\{{\bf q},{\bf k}\}])^n}_{0,k}^{(c)} .
\label{F_cumu}
\end{align}
The series can represented as Feynman diagrams leading to a systematic 
expansion in the number of loops. Up  to one-loop
order, all diagrams  are shown in
Figs.\ \ref{fig:Feynman}(c)-\ref{fig:Feynman}(i).

The single 
contributions of the Feynman diagrams shown in 
Figs.\ \ref{fig:Feynman}(c)-\ref{fig:Feynman}(i)
 to the effective energy (\ref{F_cumu}) to one-loop order 
 are 
\begin{widetext}
\begin{align}
 \frac{F'_\text{eff}[\{{\bf q}\}]_{(c)}} A &= 
  \sum_{\bf q} \frac{1}{2} \tilde f ({\bf q}) \tilde f (-{\bf q}) 
  \int_{\Lambda/b < |{\bf k}| < \Lambda} \frac{d ^2 k }{(2\pi)^2} 
  AYG\left({\bf k}+\frac{\bf q}{2}\right)
 \left[q_iP_{ij}^T\left( {\bf k} - \frac{\bf q}{2}\right)
  \left(k_j + \frac{q_j}{2}\right)\right]^2, 
  \label{diagconc}\\
 \frac{F'_\text{eff}[\{{\bf q}\}]_{(d-g)}} A &= 
  \sum_{\bf q} \frac{1}{2} \tilde f({\bf q}) \tilde f(-{\bf q})
   \int_{\Lambda/b < |{\bf k}| < \Lambda} \frac{d^2 k }{(2\pi)^2} 
 \frac {(-1)Y^2A^2}{k_B T R_0^2} 
  G\left({\bf k}+\frac{\bf q}{2}\right)G\left({\bf k}-\frac{\bf q}{2}\right)
 \nonumber\\
 &~~~~\times \bigg[
   \left[q_i P_{ij}^T\left({\bf k} + \frac{\bf q}{2}\right) 
  \left(k_j - \frac{q_j}{2} \right)\right]^2
 - \left[q_iP_{ij}^T \left({\bf k} - \frac{\bf q}{2}\right)
  \left(k_j + \frac{q_j}{2} \right)\right] 
  \left[q_i P_{ij}^T\left({\bf k} + \frac{\bf q}{2}\right) 
   \left( k_j - \frac{q_j}{2}\right)\right]
\nonumber\\
&~~~~
  + 2\left[q_iP_{ij}^T \left({\bf k} - \frac{\bf q}{2}\right)
  \left(k_j + \frac{q_j}{2} \right)\right] 
  \left[\left(k_i - \frac{q_i}{2} \right) P_{ij}^T({\bf q})
   \left( k_j + \frac{q_j}{2}\right)\right]
+ \frac{1}{2} \left[\left(k_i - \frac{q_i}{2} \right)P_{ij}^T({\bf q}) 
  \left(k_j + \frac{q_j}{2} \right)\right]^2\bigg],
\label{diagcond-g}\\
\frac{F'_\text{eff}[\{{\bf q}\}]_{(h)}} A &= 
  \sum_{\substack{{\bf q}\ne 0 \\ {\bf q}_2 + {\bf q}_3 = -{\bf q}}}
  \frac{Y}{2R_0}
  \left[q_{2i}P_{ij}^T ({\bf q}) q_{3j}\right] \tilde f({\bf q})\tilde f ({\bf
    q}_2) \tilde f ({\bf q}_3)
\nonumber\\
&~~~~~~\times \int_{\Lambda/b < |{\bf k}| < \Lambda} \frac{d^2 k}{(2\pi)^2}
 \frac {(-1)YA^2}{2k_B T} G\left({\bf k} + \frac{\bf q}{2} \right) 
   G\left({\bf k} - \frac{\bf q}{2} \right)
\left[\left(k_i - \frac{q_i}{2} \right) P_{ij}^T({\bf q}) 
  \left(k_j + \frac{q_j}{2}\right)\right]^2,
\label{diagconh}\\
\frac{F'_\text{eff}[\{{\bf q}\}]_{(i)}} A &= 
  \sum_{\substack{{\bf q}_1+{\bf q}_2 = {\bf q}\ne 0 \\ {\bf q}_3 + {\bf q}_4=
      -{\bf q}\ne 0}}
 \frac{Y}{8} [q_{1i}P_{ij}^T({\bf q}) q_{2j}][q_{3i}P_{ij}^T ({\bf q}) q_{4j}]
 \tilde f ({\bf q}_1) \tilde f({\bf q}_2) \tilde f ({\bf q}_3) \tilde f ({\bf
   q}_4)
 \nonumber\\
&~~~~~~\times \int_{\Lambda/b < |{\bf k}| < \Lambda} \frac{d^2 k }{(2\pi)^2}
 \frac{(-1)YA^2}{2k_B T} G\left({\bf k} + \frac{\bf q}{2} \right) 
  G\left( {\bf k} - \frac{\bf q}{2} \right)
\left[\left(k_i -\frac{q_i}{2} \right) P_{ij}^T ({\bf q})
   \left(k_j + \frac{q_j}{2}\right)\right]^2.
\label{diagconi}
\end{align}
\end{widetext}

The relevant terms for
the renormalization of $\kappa$, $p$, and $Y$ are terms of order $q^4$, 
$q^2$, and $q^0$ in Eqs.\ \eqref{diagconc} and \eqref{diagcond-g},
 respectively. 
They renormalize the quadratic part $F_0$ in (\ref{eq:F0}). 
The
contributions to three- and four-point vertices \eqref{diagconh} and
\eqref{diagconi} can also be used to calculate a 
 renormalization of $Y$  in the cubic and quartic part 
$F_\text{int}$ in (\ref{eq:Fint}) by considering terms $q^0$.

After performing the momentum-shell integrals over 
$\Lambda/b < |{\bf k}| < \Lambda$ in 
\eqref{diagconc} and \eqref{diagcond-g} [approximating integrals
over the momentum shell $\Lambda/b < |{\bf k}| < \Lambda$
as 
$\int_{\Lambda/b < |{\bf k}| < \Lambda} d^2 k g({\bf k}) \approx 
2\pi g(\Lambda) \Lambda s$ for scale factors  $b=e^s\approx 1-s$
with  $s\ll 1$]  and  after subsequent rescaling
according to the Eqs.\ \ref{rescale} with $b=e^s$, 
we find
new elastic parameters $\kappa'(s)$, $Y'(s)$,
and $p'(s)$, that retain the form of the free enthalpy \eqref{eq:F}
upon a change of scale by a factor  $b$.
Their RG flow  for an infinitesimal change of scale $b\approx 1-ds$ 
is described by $\beta$-functions
\begin{widetext}
\begin{subequations}
\begin{align}
 \beta_\kappa &= \frac {d\kappa'}{d s} 
  = 2(\zeta_f -1)\kappa' +
 \frac {3k_B T Y' \Lambda^2}{16\pi D} 
- \frac {3k_B T Y'^2\Lambda^2}{8\pi R_0'^2D^2}
  \left[\frac{11}{12} + \frac{I_{\kappa 1}} {D^2} 
   + \frac {I_{\kappa 2}}{D^4}\right],
\label{beta_kappa_app}\\
\beta_Y &= \frac {d Y'}{d s} 
   = 2\zeta_f Y' - \frac {3k_B T Y'^2\Lambda^6}{32\pi D^2},
\\
\beta_p &= \frac{dp'}{ds} 
   = (2\zeta_f +1)p' + \frac {3k_B T Y'^2 \Lambda^4}{4\pi R_0'^3 D^2}
     \left[1 + \frac {I_p} {D^2}\right],
\\
\beta_R &= \frac {dR_0'}{ds} = -R_0',
\end{align}
\label{BetaFunctions_app}
\end{subequations}
\end{widetext}
with  the denominator
 \begin{align}
  D &\equiv  \kappa'\Lambda^4 - \frac {p'R_0'\Lambda^2} 2 + \frac {Y'}{R_0'^2}.
\label{eq:D_app}
\end{align}
 Calculating
$\beta$-functions for $Y$ using the three- and four-point vertices
  \eqref{diagconh} or \eqref{diagconi} yields
the same $\beta_Y$, but with the terms 
$(3\zeta_f - 1)Y'$ or $(4\zeta_f -2)Y'$ instead of $2\zeta_fY'$ 
from a different rescaling. 
In order for all of these to produce the same
result, $\zeta_f = 1$ has to be  chosen.

The terms $I_{\kappa 1}$, $I_{\kappa 2}$, and $I_p$ in the RG
equations \eqref{BetaFunctions} are
\begin{widetext}
\begin{subequations}
\begin{align}
I_{\kappa 1}  &= \frac{1}{12} \left[3\frac{p'Y'}{R_0'} - p'^2R_0'^2\Lambda^4
   - 16\frac {\kappa'Y'}{R_0'^2}\Lambda^4 + 7 \kappa'p'R_0'\Lambda^6 
  - 8 \kappa'^2 \Lambda^8\right],
\label{eq:Ikappa1}\\
I_{\kappa 2} &= \frac{1}{768} \bigg[ - \frac{24Y'^3\kappa'\Lambda^4}{R_0'^6}
 + \frac {Y'^2}{R_0'^4} (9p'^2R_0'^2\Lambda^4 - 76p'R_0'\kappa'\Lambda^6 
  + 268\kappa'^2\Lambda^8)
\nonumber\\
&~~~~+\frac{Y'}{R_0'^2} ( -5p'^3R_0'^3\Lambda^6 + 52p'^2R_0'^2\kappa'\Lambda^8 
- 204p'R_0'\kappa'^2\Lambda^{10} + 160\kappa'^3\Lambda^{12})
\nonumber\\
&~~~~+ (p'^4R_0'^4\Lambda^8 - 12p'^3R_0'^3\kappa'\Lambda^{10} 
  + 56p'^2R_0'^2\kappa'^2\Lambda^{12} - 96p'R_0'\kappa'^3\Lambda^{14} 
  + 60\kappa'^4\Lambda^{16})\bigg],
\\
I_p &= \frac{1}{48} \left[\frac {Y'}{R_0'^2}(3p'R_0'\Lambda^2 
  - 16\kappa'\Lambda^4) + (-p'^2R_0'^2\Lambda^4 + 7p'R_0'\kappa'\Lambda^6 
  - 8\kappa'^2\Lambda^8)\right].
 \end{align}
\label{eq:Is}
\end{subequations}
\end{widetext}
The function $\beta_\kappa$ in 
 (\ref{BetaFunctions_app}) slightly differs from the 
results in Ref.\ \cite{Kosmrlj2017}. 
 Differences are in the two terms 
$({3 k_B T Y'^2 \Lambda^2}/{8\pi R_0^2D^2})\left(\frac{11}{12}+
 \frac{I_{\kappa 1}}{D^2}\right)$
 in Eq.\ (\ref{beta_kappa_app}) for the function $\beta_\kappa$.
 First,  the function $I_{\kappa 1}$ in (\ref{eq:Ikappa1})  differs
 from the corresponding function in Ref.\   \cite{Kosmrlj2017}; second,
 the factor $11/12$ is unity in   Ref.\   \cite{Kosmrlj2017}.
 Both differences exactly compensate each other such that 
 we have the exact same RG equations as in  Ref.\   \cite{Kosmrlj2017}.

The  scale-dependent, i.e., renormalized but unrescaled 
 quantities $\kappa_R(q)$, $p_R(q)$, and $Y_R(q)$ defined via 
Eq.\ (\ref{eq:Rquantity}) obey RG equations
\begin{subequations}
\begin{align}
 -q \frac {d \kappa_R}{d q} 
  &= 
 \frac {3k_B T Y_R q^2}{16\pi D_R} \nonumber \\ 
  &~~~- \frac {3k_B T Y_Rq^2}{8\pi R_0^2D_R^2}
  \left[ {\frac{11}{12}}  + \frac{I_{\kappa 1}} {D_R^2} 
   + \frac {I_{\kappa 2}}{D_R^4}\right],
\label{beta_kappaR}\\
 -q \frac {d Y_R}{d q} 
   &= - \frac {3k_B T Y_R^2q^6}{32\pi D_R^2},\\
 -q  \frac{\d p'}{d q} 
   &=  \frac {3k_B T Y_R^2 q^4}{4\pi R_0^3 D_R^2}
     \left[1 + \frac {I_p} {D_R^2}\right],
\end{align}
\label{BetaFunctionsR}
\end{subequations}
with  
$D_R(q) = \kappa_R (q) q^4 - 
     p_R(q) R_0 q^2/2 + {Y_R(q)}/{R_0^2}$ as in Eq.\ (\ref{eq:DRq})
and terms  $I_{\kappa 1}$,~$I_{\kappa 2}$,  and $I_p$  
which are given by Eq.\ (\ref{eq:Is}), where we similarly replace 
$\kappa' \to \kappa_R$, $Y'\to Y_R$, $p'\to p_R$, $R_0' \to R_0$, 
and $\Lambda \to q$.

\renewcommand{\theequation}{B\arabic{equation}}

\section{Energy minimization using the Surface Evolver}
\label{sec:SE}

The SURFACE EVOLVER is a program developed by Brakke for
calculating the lowest-energy shape of a triangulated surface 
 with definable  energies and
constraints \cite{SEExpMath}. 
The surface consists of vertices, which are connected by edges,
which in turn form facets. 
After specifying the  energy functional,
the minimal energy shape is found iteratively by
displacing vertices either following the steepest gradient or with more
refined conjugate gradient methods and according to the applied 
constraints.

In order to create a sphere in the SURFACE EVOLVER,  a cube is
created first. 
This surface is successively refined by dynamical triangulation
using  a 
simple liquid surface energy $E = \sigma \int d S$
   with constant surface tension $\sigma$ until 
an acceptable spherical shape is reached. 
Therefore, the resulting final triangulation contains six fourfold
disclinations corresponding to the six faces of the original cube.
Once this has been accomplished, the  
surface tension is set to zero, the triangulation is fixed,
 and the appropriate 
dimensionless elastic  energies (\ref{eq:Fdim}) are activated
(i.e., we measure energies in units of $YR_0^2$ and lengths 
in units of $R_0$ in the simulation);
the  newly created sphere is defined as the  relaxed state
of the surface in the  elastic  energies (\ref{eq:Fdim}).

The  elastic energies are then minimized for a given 
dimensionless pressure $p$ (measured in units of $R_0/Y$).
We determine the barrier for $0<p\le p_c$, where the sphere 
first compresses uniformly. Then we map out the energy landscape 
of the buckling energy barrier by selecting 
 two vertices on opposite sides of the compressed sphere of radius $R(p)<R_0$,
which are  fixed in place with a distance $z_0=2R(p)$. 
By keeping a constraint on the distance $z$ between these two vertices and 
decreasing this  distance $z$ starting from $z_0$, 
we control the size of the 
dimple. A similar procedure has been used in Ref.\ \cite{Paulose2012} 
for fluctuating spherical shells. 
We obtain the enthalpy $F=F(z)$ as a function of $z$  
for a given pressure $p$ (see Fig.\ \ref{fig:Fz})
  and can determine its maximum $F_\text{max}$.
The energy barrier  is $F_\text{B} = F_\text{max}- F(z=z_0)$ with $z_0=2R(p)$.
By changing the pressure $p$, we can numerically determine the energy barrier 
as a function of pressure $F_\text{B} = F_\text{B}(p)$.

\begin{figure}
\begin{center}
\includegraphics[width=1\linewidth]{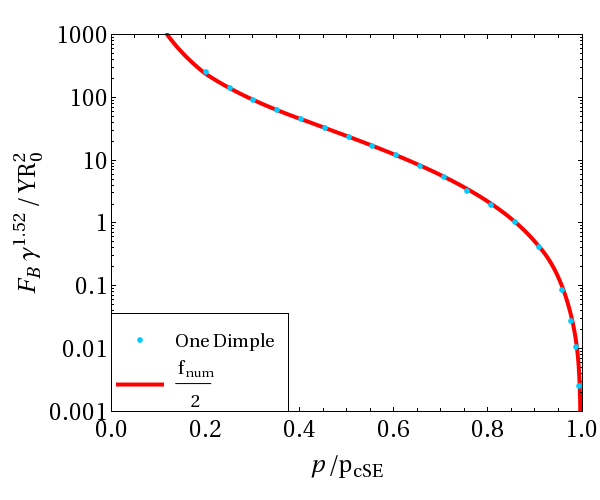} 
\caption{
 Energy barrier as a function of pressure
for one dimple  (light blue points) and
  half of the approximated energy barrier for two dimples 
(solid dark red line) from 
numerical simulations;
   see Eq.\ (\ref{eq:FBarrnum})
(for $\gamma=10^4$ and $\nu=0.3$).
}
\label{fig:2dimple}
\end{center}
\end{figure}

Constraining the distance $z$ between two opposite points on the sphere
generates two dimples upon decreasing $z$;
if one of the points is fixed before decreasing $z$, only a single 
dimple is created. 
For sufficiently small dimples, the energy barrier for a sphere that forms
only one dimple is half of the energy barrier for a sphere with two dimples
\cite{Hutchinson2017}, because the interaction of two small 
dimples on opposite
sides of a sphere can be neglected; see Fig.\ \ref{fig:2dimple}. 

\subsection{Discretization effects}
\label{sec:discrete}

\begin{figure*}
\begin{center}
\includegraphics[width=1\linewidth]{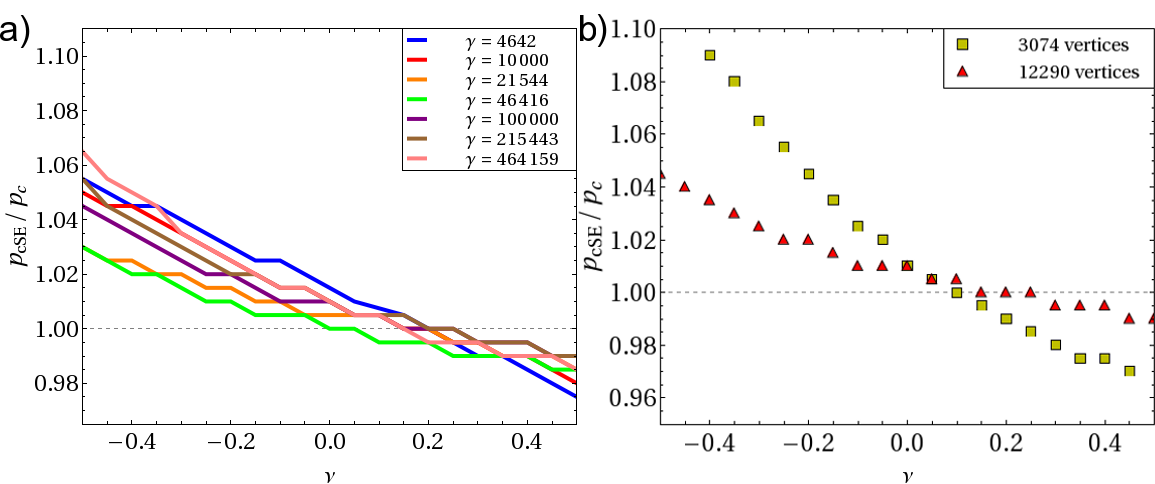}
\caption{
  Approximate critical buckling pressure $p_\text{c,SE}$ 
  as determined with the SURFACE EVOLVER 
as a
  function of Poisson's number $\nu$  (a)
   for different values of $\gamma$ with 12290 vertices
   and (b) for different vertex counts at $\gamma=100000$. For
  $\nu\approx 0.15$, $p_\text{c,SE}$  roughly agrees with
  $p_c$. Therefore, and 
  because the estimate $p_\text{c,SE}$
   for the critical buckling pressure becomes
  more inaccurate for larger $\nu$ (as the SURFACE EVOLVER does not
  reach the buckled shape in a reasonable amount of computation 
  time even for $p>p_\text{c,SE}$), 
  Poisson ratios $\nu=0.05...0.5$ are used
  for the numerical calculations. 
  For larger vertex counts, the approximate
  critical buckling pressure moves toward $p_c$, indicating that the
  deviation is mostly a triangulation effect. 
  However, disclinations
  most likely also play a role. A vertex count of 12290 was used
  for all remaining numerical  simulations.
}
\label{fig:p_cSEapp}
\end{center}
\end{figure*}

In simulating buckling, discretization effects occur for the 
 triangulated surfaces in the SURFACE EVOLVER. 
In the SURFACE EVOLVER, buckling does not occur exactly at the 
 classical buckling pressure $p_{c} = 4 (Y/R_0)\gamma^{-1/2}$ but 
at an approximate  value $p_\text{c,SE}$, which deviates 
because of discretization effects. 
This value is determined numerically 
 by gradually increasing the pressure
 until the average of all 
vertex distances to the sphere's
 center becomes smaller than $0.9\, R_0$, 
at which point buckling has occurred.
This approximated buckling pressure
$p_\text{c,SE}$ is shown in Fig.\ \ref{fig:p_cSEapp} for different 
Poisson ratios $\nu$ and different 
triangulations. 
In order to obtain an energy barrier $F_\text{B}(p/p_c)$, which 
properly vanishes at $p=p_c$, i.e.,  $F_\text{B}(1) = 0$, we use
 the numerically determined buckling pressure $p_\text{c,SE}$
instead of $p_c$. 
We analyze the numerical results for the energy barrier 
in Figs.\ \ref{fig:gamma-E}, \ref{fig:nu-iso}, and \ref{fig:Num-Pog} 
also as a function of $p/p_\text{c,SE}$ rather than $p/p_c$.

\subsection{Numerical approximation for the energy barrier}
\label{FBarrNum}

A very accurate  approximation for the 
numerically determined energy barrier 
of two dimples can be given as 
\begin{align}
 \frac {2F_\text{B,num}(p/p_c, \nu, \gamma)}{YR_0^2}
   &=  2f_\nu(\nu)\gamma^{-1.52}f_\text{p,num}(p/p_c)
\label{eq:FBarrnum}
\end{align}
with a scaling function 
\begin{align}
   2f_\text{p,num}(x) &= 
\begin{cases}
f_1(x) & x>0.910\\
f_2(x) & 0.910>x>0.418\\
f_3(x) & 0.418>x>0.198\\
f_4(x) & 0.198>x                     
\end{cases}
\end{align}
with
\begin{widetext}
\begin{align}
f_1(x) &= -8490.551 - 12199.07 \exp(-0.07155726 x) + 4504.429/x^6 
 \nonumber\\
 &~~
  - 10596.10/x^5- 619.6270/x^4 + 9590.704/x^3 + 7692.978/x^2 
\nonumber\\
 &~~
  - 2309.406/x + 1804.463 x + 18542.43 x^2 - 8762.693 x^3,
\nonumber\\
f_2(x) &= -74.00287 - 691.7514 \exp(-1.026248 x) + 0.01542655/x^6 
\nonumber\\
 &~~ 
 +  0.005939230/x^5- 0.5782996/x^4 + 12.10501/x^3 - 83.49014/x^2 \nonumber\\
 &~~ 
  + 389.3926/x + 6.859172 x - 0.7489285 x^2 - 1.683589 x^3,
\nonumber\\
f_3(x) &= -9.090486 + 12.55716/x^{2.272675},
\nonumber\\
f_4(x) &= 20.05584 + 4.911410/x^{2.815042},
\label{eq:fnum}
\end{align}
\end{widetext}
and
\begin{align}
f_\nu(x) &= 0.9754744+0.08569536 \nu.
\label{eq:fnu}
\end{align}
The $\nu$ dependence is very weak.

\begin{figure}
\begin{center}
\includegraphics[width=1\linewidth]{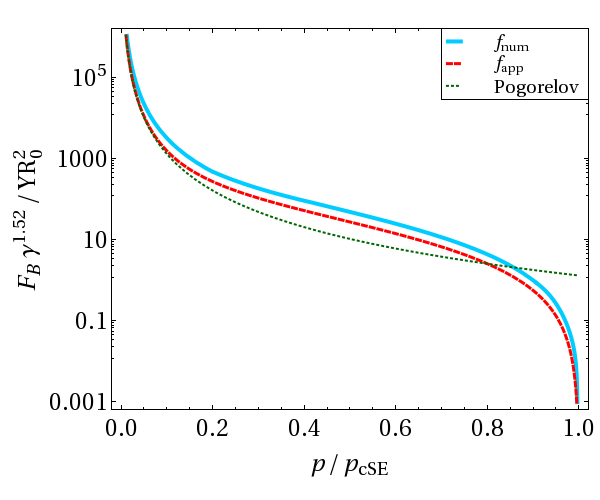} 
\caption{
 Energy barrier for a sphere with one dimple 
as a function of pressure
(for $\gamma=10^4$ and $\nu=0.3$)
as  obtained from  numerical
  simulations [Eq.\ (\ref{eq:FBarrnum}), light blue solid line], 
  from the  Pogorelov model [Eq.\ (\ref{FbarrTh}), green dotted line], 
and according to the approximation (\ref{eq:FBarrapp}) 
(dark red dashed line).
}
\label{fig:Num-Pog}
\end{center}
\end{figure}

A simpler approximation formula, which is motivated by the Pogorelov 
result (\ref{FbarrTh}), is 
\begin{align}
 \frac {F_\text{B,num}(p/p_c, \gamma)}{YR_0^2}
   &= \gamma^{-3/2}f_\text{p,app}(p/p_c) ~~\text{with}
\nonumber\\
   f_\text{p,app}(x) &= 1.44\,(1-x)^2 (x^{-3} +34.1\,x^{-1}).
\label{eq:FBarrapp}
\end{align}
This simple approximation agrees within $20\%$ with Eq.\ (\ref{eq:FBarrnum})
(see Fig.\ \ref{fig:Num-Pog}). 

\bibliography{references}

\end{document}